\documentclass{aa}  
\usepackage{graphicx}
\usepackage{txfonts}
\usepackage{float}
\usepackage{subfig}
\usepackage{xcolor}
\defcitealias{Quirola2022}{Paper~I}
\usepackage[draft]{hyperref}

\begin{document} 

   \titlerunning{On a Magnetar Origin of FXTs}
   \authorrunning{Quirola-V\'asquez et al.}

   \title{Probing a Magnetar Origin for the population of Extragalactic Fast X-ray Transients detected by \emph{Chandra}}
   
   \author{J. Quirola-V\'asquez
          \inst{1,2,3},
          F. E. Bauer\inst{1,2,4},
          P.~G. Jonker\inst{5,6},
          W. N. Brandt\inst{7,8,9},
          D. Eappachen\inst{6,5},
          A. J. Levan\inst{5,10},
          E. Lopez\inst{3},
          B. Luo\inst{11,12}
          M. E. Ravasio\inst{5,13},
          H. Sun\inst{14},
          Y. Q. Xue\inst{15,16},
          G. Yang\inst{17,18},
          \and
          X. C. Zheng\inst{19}
          }
   \institute{Instituto de Astrof\'isica, Pontificia Universidad Cat\'olica de Chile, Casilla 306, Santiago 22, Chile\\
              \email{jaquirola@uc.cl}
         \and
             Millennium Institute of Astrophysics (MAS), Nuncio Monse$\tilde{\rm n}$or S\'otero Sanz 100, Providencia, Santiago, Chile
        \and
             Observatorio Astron\'omico de Quito, Escuela Polit\'ecnica Nacional, 170136, Quito, Ecuador
        \and
            Space Science Institute, 4750 Walnut Street, Suite 205, Boulder, Colorado 80301, USA
        \and
            Department of Astrophysics/IMAPP, Radboud University, P.O. Box 9010, 6500 GL, Nijmegen, The Netherlands
        \and
            SRON Netherlands Institute for Space Research, Niels Bohrweg 4, 2333 CA Leiden, The Netherlands
        \and
            Department of Astronomy \& Astrophysics, 525 Davey Laboratory, The Pennsylvania State University, University Park, PA 16802, USA
        \and
            Institute for Gravitation and the Cosmos, The Pennsylvania State University, University Park, PA 16802, USA
        \and
            Department of Physics, 104 Davey Laboratory, The Pennsylvania State University, University Park, PA 16802, USA
        \and
            Department of Physics, University of Warwick, Coventry, CV4 7AL, UK
        \and
            School of Astronomy and Space Science, Nanjing University
        \and
            Key Laboratory of Modern Astronomy and Astrophysics (Nanjing University), Ministry of Education, Nanjing 210093, China
        \and
            INAF - Osservatorio Astronomico di Brera, via E. Bianchi 46, 23807, Merate (LC), Italy
        \and
            National Astronomical Observatories, Chinese Academy of Sciences, Beijing 100101, People’s Republic of China
        \and
            CAS Key Laboratory for Research in Galaxies and Cosmology, Department of Astronomy, University of Science and Technology of China, Hefei 230026, China
        \and
            School of Astronomy and Space Science, University of Science and Technology of China, Hefei 230026, China
        \and
            Kapteyn Astronomical Institute, University of Groningen, P.O. Box 800, 9700 AV Groningen, The Netherlands
        \and
            SRON Netherlands Institute for Space Research, Postbus 800, 9700 AV Groningen, The Netherlands
        \and
            Leiden Observatory, Leiden University, PO Box 9513, NL-2300 RA, Leiden, the Netherlands
             }

   \date{Received August 1, 2023; accepted December 28, 2023}

\abstract
{Twenty-two extragalactic fast X-ray transients (FXTs) have now been discovered from two decades of \emph{Chandra} data (analyzing ${\sim}$259~Ms of data), with 17 associated with distant galaxies (${\gtrsim}$100~Mpc). Different mechanisms and progenitors have been proposed to explain their properties; nevertheless, after analyzing their timing, spectral parameters, host-galaxy properties, luminosity function, and volumetric rates, their nature remains uncertain.}
{We interpret a sub-sample of nine FXTs that show a plateau or a fast-rise light curve within the framework of a binary neutron star (BNS) merger magnetar model.}
{We fit their light curves and derive magnetar (magnetic field and initial rotational period) and ejecta (ejecta mass and opacity) parameters. This model predicts two zones: an orientation-dependent free zone (where the magnetar spin-down X-ray photons escape freely to the observer) and a trapped zone (where the X-ray photons are initially obscured and only escape freely once the ejecta material becomes optically thin). We argue that six FXTs show properties consistent with the free zone and three FXTs with the trapped zone.}
{This sub-sample of FXTs has a similar distribution of magnetic fields and initial rotation periods to those inferred for short gamma-ray bursts (SGRBs), suggesting a possible association.
We compare the predicted ejecta emission fed by the magnetar emission (called merger-nova) to the optical and near-infrared upper limits of two FXTs, XRT~141001 and XRT~210423 where contemporaneous optical observations are available. The non-detections place lower limits on the redshifts of XRT~141001 and XRT~210423 of $z{\gtrsim}1.5$ and ${\gtrsim}0.1$, respectively.}
{If the magnetar remnants lose energy via gravitational waves (GWs), it should be possible to detect similar objects with the current advanced LIGO detectors out to a redshift $z{\lesssim}0.03$, while future GW detectors will be able to detect them out to $z{\approx}0.5$.}

    
   \maketitle



\section{Introduction} \label{sec:intro}

From their first three observational runs, the LIGO/VIRGO gravitational wave (GW) detectors discovered 90 binary black holes (BBHs), two binary neutron stars (BNS), and two binary neutron star-black hole (NS-BH) systems \citep{Nitz2023}. Among these, the unique discovery of GW~170817 \citep{Abbott2017e} and its associated broadband electromagnetic (EM) detections \citep{Abbott2017a,Abbott2017b} marked the arrival of the multi-messenger astronomy era with the detection of the short gamma-ray burst (SGRB) GRB~170817A \citep{Goldstein2017,Savchenko2017,Abbott2017b}, the optical and near-infrared (NIR) kilonova AT2017gfo \citep{Arcavi2017,Lipunov2017,Coulter2017,Tanvir2017,Soares2017,Valenti2017,Abbott2017c}, and the broadband radio-to-X-ray afterglow \citep{Alexander2017,Evans2017,Hallinan2017,Margutti2017,Troja2017,Savchenko2017,Abbott2017b,Sugita2018}, which collectively reinforced related theoretical models and provided novel parameter constraints \citep[e.g.,][]{Li1998,Metzger2010,Goriely2011,Roberts2011,Kasen2013,Fernandez2016,Smartt2017,Metzger2019}.

The merger remnant of an NS–BH binary is expected to be a BH, while four different remnant possibilities have been argued to result from a BNS merger: 
a BH; 
a differential-rotation-supported hypermassive NS lasting ${\sim}$30--300~ms before collapsing to a
BH \citep[hereafter HMNS;][]{Sun2017,Margalit2017,Pooley2018,Rezzolla2018,Ruiz2018,Margalit2019}; 
a rigid-rotation-supported supra-massive neutron star (with a low dipolar magnetic field; hereafter SMNS) that lasts for tens of seconds to ${\gtrsim}10^4$~s \citep{Margalit2019} before collapsing to a BH \citep{Sun2017,Ai2018};
and a stable NS that persists.

The outcome after the BNS merger depends on the total system mass \citep[typically ${\approx}$2.5--2.7~$M_\odot$ among Galactic BNS;][]{Kiziltan2013,Martinez2015}, and the unknown NS equation of state \citep[EoS;][]{Sun2017}. 
Indeed, recently, some massive NSs have been identified, for instance, PSR~J0952-0607 \citep[$M_{\rm NS}{=}2.35\pm0.17$~$M_\odot$;][]{Romani2022} and PSR~J0740+6620 \citep[$M_{\rm NS}{=}2.08\pm0.07$~$M_\odot$;][]{Fonseca2021}, providing the most severe constraints on the dense-matter EoS.
The possibility of BNS mergers producing a long-lived NS has been suggested in the literature to interpret some of the X-ray features observed in SGRB afterglows, including X-ray flares \citep{Dai2006,Gao2006}, GRB extended emission \citep{Metzger2008,Gompertz2014}, and the so-called internal plateau\footnote{An ``internal plateau'' is a plateau followed by a quick decay (power-law index of ${\lesssim}-3$), which is not explained by the evolution of the synchrotron emission from a decelerating forward shock, but could be interpreted within the central engine framework \citep[e.g.,][]{Zhang2001,Tang2019}.} 
observed in a good fraction of SGRBs \citep{Rowlinson2010,Rowlinson2013,Lu2015,Gao2016,Li2016,Li2016a,Yu2018,Piro2019}. This interpretation has allowed constraints on the maximum NS remnant mass of ${\lesssim}$2.5--2.8~$M_\odot$ \citep{Lasky2014,Gao2016,Lu2015,Lu2017}.

A potentially related phenomenon with BNS merger is extragalactic Fast X-ray transients (FXTs), short flashes of X-ray photons with durations from a few minutes to hours, which have been observed in the $\sim$0.3--10~keV X-ray band by \emph{Chandra}, \emph{XMM-Newton} and \emph{Swift}-XRT \citep[e.g.,][]{Soderberg2008,Jonker2013,Glennie2015,Irwin2016,Bauer2017,Lin2018,Lin2019,Xue2019,Alp2020,Novara2020,Lin2020,Ide2020,Pastor2020,Lin2021,Lin2022,Eappachen2023,Quirola2022,Quirola2023}.
At present, their nature remains poorly understood in part due to the lack of 
EM counterparts at other wavebands. 
Critically, while 34 extragalactic FXTs have been identified to date, both serendipitously and through careful searches, only in one case, XRT~080109/SN~2008D \citep{Soderberg2008,Mazzali2008,Modjaz2009}, has there been a detection of a multi-wavelength counterpart after the outburst. The others have only been identified via archival data-mining from days \citep[e.g.,][]{Luo2014,Lin2021} to years after the outbursts \citep[e.g.,][]{Alp2020,Lin2020,Lin2022,Quirola2022,Quirola2023}. Moreover, only a fraction of FXTs has clear host-galaxy associations \citep[e.g., 13 FXTs detected by \emph{Chandra} have been associated with extended objects;][]{Quirola2022,Quirola2023,Eappachen2022,Eappachen2023}, while far fewer have firm distance and energetic constraints \citep[e.g.,][]{Soderberg2008,Irwin2016,Bauer2017,Xue2019,Alp2020,Novara2020,Lin2022,Eappachen2022,Eappachen2023,Quirola2022,Quirola2023}.

A variety of theoretical models have been proposed for the origin of extragalactic FXTs, such as: 
$i)$ X-ray emission produced from the shock breakout (SBO; $L_{\rm X,peak}{\sim}$10$^{42}$--10$^{44}$~erg~s$^{-1}$) of core-collapse supernova (CC-SN) once it crosses the stellar material \citep[e.g.,][]{Soderberg2008,Nakar2010,Waxman2017,Novara2020, Alp2020};
$ii)$ the X-ray emission from the off-axis long GRBs ($L_{\rm X,peak}{\lesssim}$10$^{45}$~erg~s$^{-1}$) is produced by a wider, mildly relativistic cocoon jet once it breaks the surface of a massive progenitor star \citep{Ramirez2002,Zhang2004b,Nakar2015,Zhang_book_2018,Delia2018};
$iii)$ tidal disruption events (TDEs; $L_{\rm X,peak}{\sim}$10$^{42}$--10$^{50}$~erg~s$^{-1}$) 
involving a white dwarf (WD) and an intermediate-mass black hole (IMBH), whereby X-ray emission is generated from the accretion of part of the WD by the IMBH \citep[e.g.,][]{Jonker2013,MacLeod2014,Saxton2021,Maguire2020}; and
$iv)$ BNS merger \citep[$L_{\rm X,peak}{\sim}$10$^{44}$--10$^{51}$~erg~s$^{-1}$, e.g.,][]{Dai2018,Fong2015,Sun2017,Bauer2017,Xue2019}, where the X-ray photons are created by the accretion of fallback material onto the remnant BH, a wider and mildly relativistic cocoon, or the spin-down magnetar emission \citep[e.g.,][]{Metzger2014,Sun2017,Sun2019,Metzger2018b}.

Twenty-two of the above FXTs come from the analysis of \citet{Quirola2022} and \citet{Quirola2023} (hereafter Papers~I and II, respectively) using two decades of \emph{Chandra} data (Paper~I was based on the available Chandra Source Catalog 2.0 covering until the end of 2014, while Paper~II carried out a systematic reduction of \emph{Chandra} data from 2014 to April 2022). Two relatively distinct sets of FXTs were distinguished according to their distances: nearby and distant FXTs, with luminosity distances $d_L{\lesssim}$100 and $d_L{\gtrsim}$100~Mpc, respectively. Based on host-galaxy associations and estimated redshifts, an SN SBO scenario has been ruled out for many distant FXTs, based on inferred X-ray luminosity peak and isotropic energy arguments \citep{Eappachen2022,Eappachen2023,Quirola2022,Quirola2023}. The energetics and host-galaxy properties of the most distant FXTs remain consistent with an origin as a GRB or an IMBH-WD TDE. 

It has been predicted that some FXTs could be EM counterparts of BNS mergers \citep{Zhang2013,Metzger2014,Sun2017,Sun2019}. \citet{Luo2014} and \citet{Zheng2017} identified two new unusual FXTs in the 7~Ms \emph{Chandra} Deep Field-South (CDF-S) dataset, XRT~141001 and XRT~150322, denoted as ``CDF-S~XT1'' and ``CDF-S~XT2'', respectively. These two FXTs were later studied in detail by \citet{Bauer2017} and \citet{Xue2019}, respectively, and re-identified as FXTs~14 and 16 in Papers~I and II, respectively. XRT~150322 is particularly intriguing because it exhibits a flat, extended X-ray light curve that suggests a magnetar-wind origin \citep{Sun2019,Xiao2019,Lu2019}, similar to some SGRB X-ray afterglows \citep[e.g.,][]{Rowlinson2010,Rowlinson2013,Troja2019}. Its host-galaxy properties are consistent with the observed host galaxies of SGRBs \citep{Xue2019}. XRT~141001, on the other hand, is consistent with an ``orphan'' X-ray afterglow from an off-axis SGRB with weak optical emission \citep{Bauer2017,Sarin2021}. \citet{Sun2019} proposed a unified model to interpret both transients within the framework of the BNS merger magnetar model by considering different observer viewing angles (following on from earlier models developed by \citealt{Yu2013} and \citealt{Sun2017}). According to this new work, CDF-S~XT2/FXT~16 is observed from a region where the magnetar spin-down powered X-ray emission escapes freely, whereas CDF-S~XT1/FXT~14 originates from a region that is initially opaque to the X-ray emission but becomes optically thin after the ejecta expands and becomes ionised. 

In this paper, we explore a possible association of FXTs with massive, rapidly spinning magnetars produced by a BNS merger observed at different viewing angles.
The manuscript is organized as follows: we explain the magnetar model in Sect.~\ref{sec:theoretical}; we describe the data considered in this work and the fitting process in Sects.~ \ref{sec:data} and \ref{sec:fitting}, respectively;  Sects.~\ref{sec:magnetar_result}, \ref{sec:mergernova_FXTs} and \ref{sec:GW_emission} present the light curve fitting results, the upper limit constraints using optical data, and the potential for GW detections, respectively; and finally, Sect.~\ref{sec:conclusion} presents a discussion and conclusions. Throughout the paper, a concordance cosmology with parameters $H_0{=}$70~km~s$^{-1}$~Mpc$^{-1}$, $\Omega_M{=}$0.30, and $\Omega_\Lambda{=}$0.70 is adopted. Magnitudes are quoted in the AB system.

\section{Magnetar model in a nutshell} \label{sec:theoretical}

Some GRBs might have been with millisecond magnetars \citep[see for instance,][]{Rowlinson2010,Rowlinson2013,Gompertz2014,Dainotti2017}. The BNS merger model acknowledges the potential rapid formation of a BH, whose accretion could power a SGRB and its subsequent afterglow. However, the discovery of NSs with masses around ${\approx}2$~$M_\odot$ \citep{Demorest2010,Antoniadis2013,Romani2022} implies that certain mergers might result in the creation of a transient or stable rapidly-spinning magnetar instead \citep{Duncan1992,Nousek2006,Zhang2006,Metzger2008}.
As such, once the BNS merger occurs, a magnetar may be formed with enough rotational energy to prevent gravitational collapse and the direct formation of a BH \citep{Zhang2001}. Its rotational energy could be lost via GWs (via the quadrupole moment in the mass distribution) and EM radiation (via the magnetic field dipole distribution), causing the magnetar to spin down. At a certain moment, it may reach a critical point in time (from tens of seconds to ${>}10^4$~s) at which centrifugal forces can no longer support the degeneracy pressure against gravity, resulting in the formation of a BH \citep{Siegel2016a,Siegel2016b,Sun2017}. This scenario can lead to a brief X-ray plateau, as observed in certain SGRB afterglows, resulting from the injection of the spin-down energy of the magnetar and the subsequent collapse into a BH \citep{Dai1998,Fan2006,Rowlinson2010,Rowlinson2013,Lasky2017}. An alternative explanation involving fall-back accretion has also been proposed \citep{Rosswog2007}.


Here, we consider the formalism from \citet{Yu2013} and \citet{Sun2017} to describe the physical model of the magnetar emission.
Approximately isotropic X-ray emission may be produced by the internal dissipation of the magnetar wind (i.e., covering a wide solid angle); however, this process is not well understood \citep[e.g.,][]{Metzger2014}. Their luminosity tracks the dipole spin-down luminosity of the magnetar with a certain efficiency. The wind energy injection dissipated from the magnetar would produce a plateau in the X-ray light curve \citep{Zhang2001} followed by a power-law decay (like $F_X \propto t^{\alpha}$ where $\alpha{\sim}-1$ to $-2$) due to the magnetar spin down.

We focus on the scenario whereby the BNS merger produces a stable or supra-massive millisecond magnetar, which can be characterized by the initial rotational period ($P_i$) and magnetic field. In addition to the formation of the magnetar, a fraction of unbound mass is ejected (hereafter, ejecta material, $M_{\rm ej}$) due to processes that occur on a dynamical timescale \citep[see, e.g.,][]{Oechslin2007,Bauswein2013,Hotokezaka2013}, and depend on the total binary mass, mass ratio, and EoS \citep[see][and references therein]{Metzger2019}. This ejecta material, rich in $r$-process elements, powers a thermal transient called a kilonova \citep[KN;][]{Li1998}. The total dynamical ejecta mass typically ranges from $10^{-4}$--$10^{-2}$~$M_\odot$ for BNS mergers \citep[e.g.,][]{Kyutoku2013,Kyutoku2015,Foucart2017} and moves outward with velocities of 0.1--0.3$c$ \citep[some authors propose that outflow velocities greater than ${\sim}0.6c$ can occur, which through the decay of free neutrons in the outermost layers of the ejecta or prompt shock heating of the ejecta by a relativistic outflow can generate UV precursors of KNe;][]{Dean2021,Metzger2019}.


\begin{figure*}
    \centering
    \includegraphics[scale=0.6]{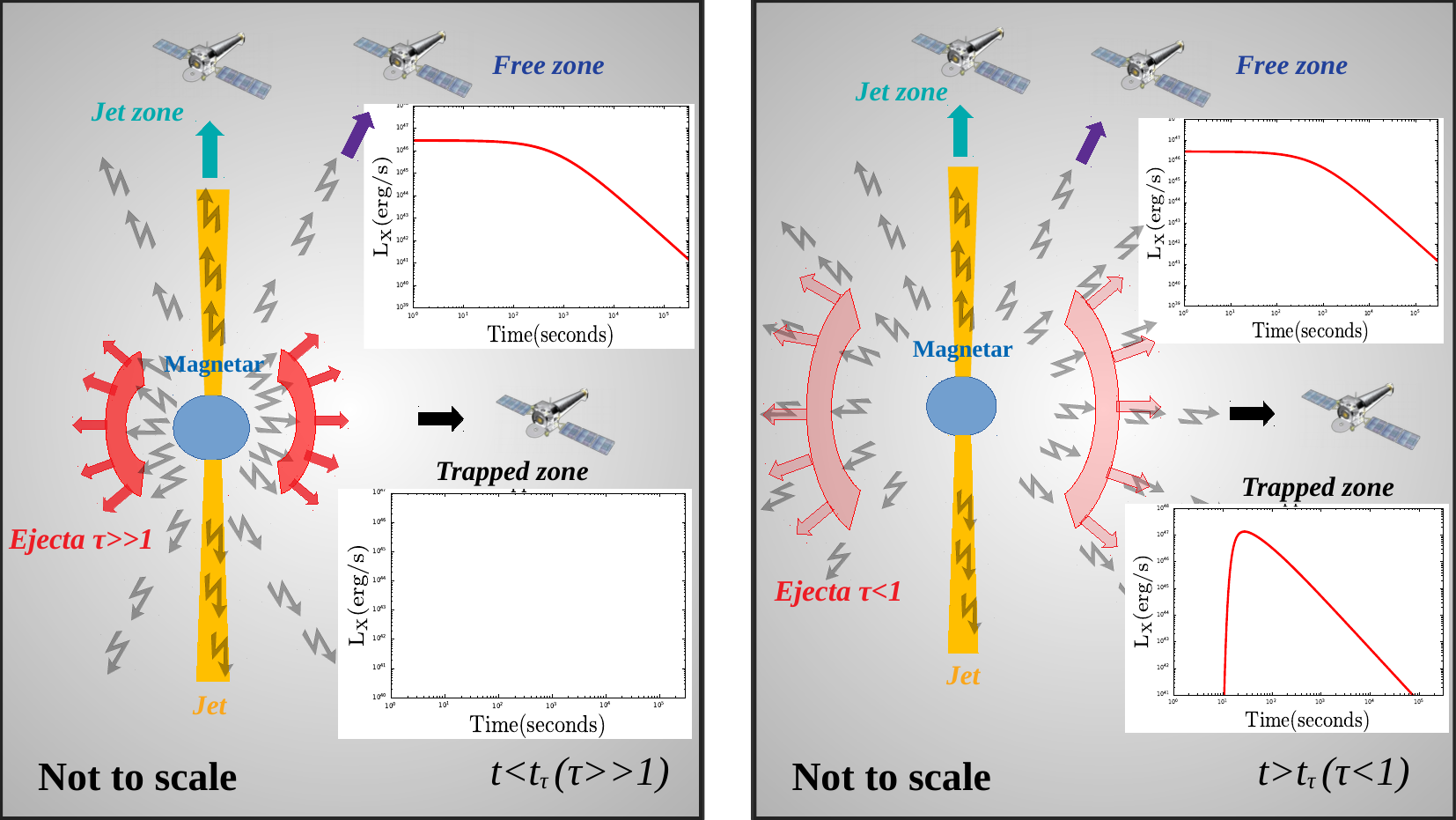}
    \caption{Cartoon illustrating the jet, free, and trapped zones, according to the magnetar model for FXTs. \emph{Left panel:} situation at time $t{<}t_\tau$. \emph{Right panel:} situation at time $t{>}t_\tau$. The insets in both panels show the light curves expected for the free and trapped zones before and after $t_\tau$.}
    \label{fig:cartoon}
\end{figure*}

Once the merger occurs, a strongly collimated, relativistic jet might be emitted perpendicular to the orbital plane of the system \citep[e.g.,][]{Narayan1992,Metzger2008,Zhang2010,Bucciantini2012,Berger2014}, and three distinct geometrical zones can be defined (see Fig.~\ref{fig:cartoon}). The magnetar emission and the ejecta material produce a configuration where the observed phenomenon depends on the observer viewing angle relative to the outflow or jet axis \citep[see][]{Yuan2015,Sun2017}:
\begin{enumerate}
    \item The {\it jet zone} is the on-axis SGRB jet direction. In this direction, the X-ray emission from the magnetar can be observed as a plateau in the SGRB afterglow \citep[e.g.,][]{Rowlinson2010,Rowlinson2013,Lu2015,Gompertz2014};
    
    \item The {\it free zone} includes any directions where the ejecta material does not obscure the magnetar emission region from the observer's line of sight. For instance, along slightly off-axis directions with respect to the jet, resulting in no strong SGRB signal, although the X-ray magnetar emission can escape freely. There could still be a weak (GRB~170817A-like) GRB along this viewing direction, which clears a funnel \citep[such a configuration could be produced by a structured jet-cocoon geometry;][]{Lazzati2017,Lamb2017} to allow X-rays to escape, but such an SGRB-like GRB~170817A would not be detectable with current $\gamma$-ray detectors beyond ${\gtrsim}$80~Mpc \citep{Zhang2018,Sun2019}.
    
    \item The {\it trapped zone} occurs from directions where the dynamical ejecta obscures the magnetar X-ray emission. In this case, the ejecta material is heated and accelerated (via the Poynting flux) by the magnetar emission, which eventually causes a magnetar-enhanced KN \citep[e.g.,][and references therein]{Yu2013,Sun2017,Metzger2014,Metzger2019}. In this case, the time when the ejecta transforms from optically thick to thin (i.e., when $t{=}t_\tau$; where $t_\tau$ denotes the time when the optical depth of the ejecta material is one) plays an important role in the dynamics of the ejecta and eventually obscured the X-ray emission.
\end{enumerate}
Below, in Sects.~\ref{sec:free_zone} and \ref{sec:trapped_zone}, we explain the mechanisms behind the free and trapped zones, respectively. In Appendix~\ref{sec:theoretical_detail}, one can find the mathematical details of this model.

\subsection{Free-zone emission} \label{sec:free_zone}

During the spin-down process, a continuous isotropic Poynting-flux-dominated outflow is launched, which could potentially inject energy and modify the ejecta material, and the afterglow properties \citep{Sun2017,Metzger2018,Ai2018,Yu2018,Piro2019,Troja2020}. The millisecond magnetar loses its rotational energy ($E_{\rm rot}$, see Eq.~\ref{eq:001}) through EM and GW radiation \citep[from non-axisymmetric deformation][]{Shapiro1983,Usov1992,Zhang2001,Gao2016,Lasky2016,Sun2017}. The evolution of the NS angular frequency considers the EM and GW losses (see Eq.~\ref{eq:002}), which both play a role in the spin-down evolution, and we define the dipole spin-down luminosity as \citep{Zhang2001,Metzger2014,Lasky2016,Sun2017,Zhang_book_2018}:
\begin{equation}
    L_{\rm sd}(t)=\frac{B_P^2R_{\rm M}^6\Omega(t)^4}{6c^3},
    \label{eq:003}
\end{equation}
where $\Omega(t)$ is the angular frequency evolution ($\Omega(t)=2\pi/P(t)$, and $P(t)$ is the rotational period), $B_{\rm P}$ is the dipole component of the magnetic field at the poles\footnote{We are assuming a pure dipole magnetic field. Nevertheless, it is known that neither purely poloidal nor toroidal \citep[these fields are responsible for causing surface currents in magnetars;][]{Soldateschi2021} magnetic topologies are stable.}, $R_{\rm M}$ is the magnetar radius, and $c$ is the speed of light. 
Importantly, a strong magnetic field provides a mechanism for extracting rotational energy from the magnetar via EM spin-down. Indeed, magneto-hydrodynamic simulations show that the magnetic fields in a BNS merger are amplified to values exceeding the field of Galactic magnetars \citep[i.e., $B_P{\sim}10^{15}-10^{16}$~G;][]{Price2006,Zrake2013,Kiuchi2014}.

\begin{figure*}
    \centering
    \hspace{-0.65 cm}
    \includegraphics[scale=0.75]{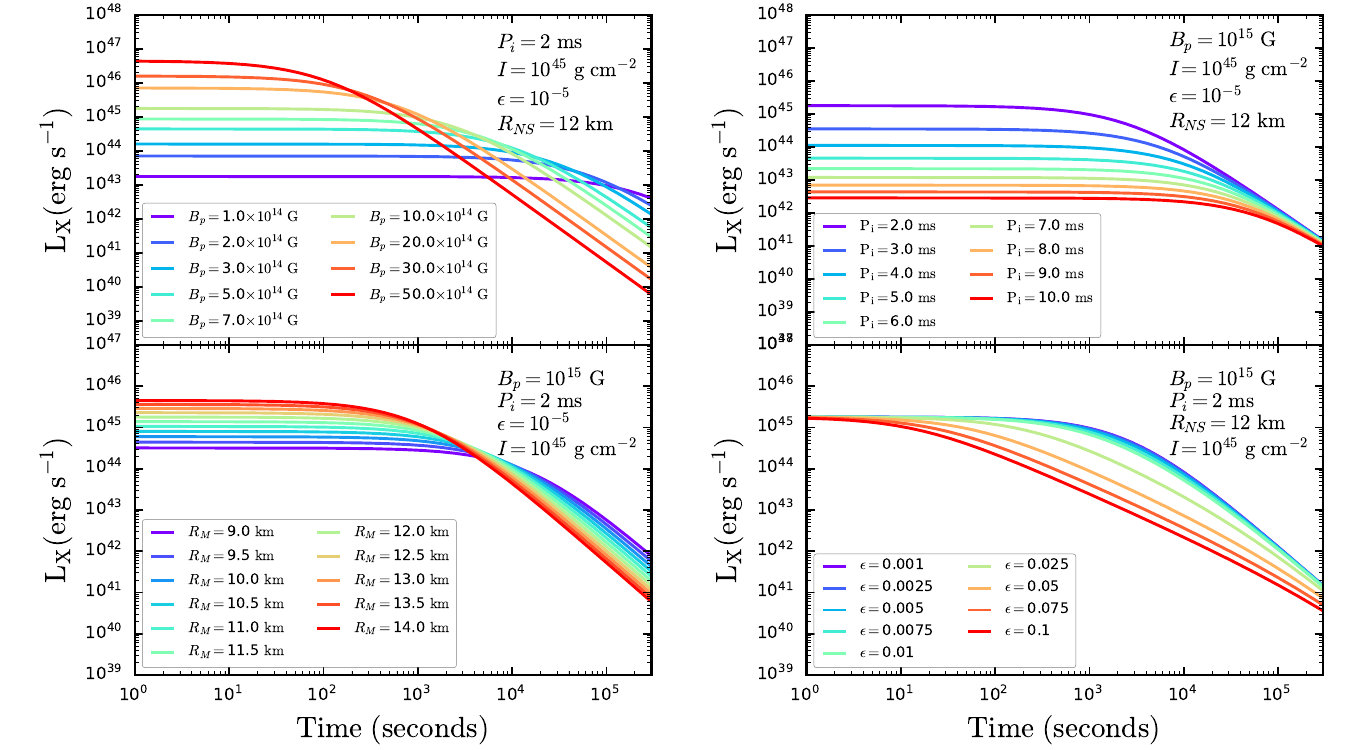}
    \caption{Theoretical light curves in the free-zone using Eqs.~\ref{eq:003}, \ref{eq:010} and \ref{eq:002}. We explore how the light curve changes as we vary the magnetic field (top-left panel), the initial spin period (top-right panel), the magnetar radius (bottom-left panel), and the ellipticity (bottom-right panel). The text in the upper right of each panel shows the fixed parameters in the numerical solutions, while the legends and colored curves indicate the varied parameters and the range in the values.}
    \label{fig:flux_comparison_free_zone}
\end{figure*}

The internal dissipation of the magnetar emission is not perfectly efficient, that is, the magnetar wind emits an X-ray luminosity that is only a fraction of $L_{\rm sd}$ (labeled as $\eta$), converting the dipole spin-down luminosity to the observed X-ray luminosity ($\eta L_{\rm sd}$). In the free zone, without obscuring material, the magnetar wind emission should be
\begin{equation}
    L_{\rm X,free}(t)=\eta L_{\rm sd}=\frac{\eta B_p^2R_{\rm M}^6\Omega^4(t)}{6c^3}.
    \label{eq:010}
\end{equation}
Typically, $\eta$ is assumed to take values of ${\approx}10^{-3}$ and to be constant in time \citep{Sun2019}. According to the geometrical/orientation-based structure (see Fig.~\ref{fig:cartoon}), the free-zone X-ray emission will be visible regardless of the evolution and dynamics of the ejecta material.

Figure~\ref{fig:flux_comparison_free_zone} shows the numerical solution for the free zone (using Eqs.~\ref{eq:003}, \ref{eq:010} and \ref{eq:002}, and under the assumption of $\eta{=}$10$^{-3}$) exploring the parameter space, that is, varying the magnetic field component ($B_{\rm P}$), the initial rotational period ($P_i$), magnetar radius ($R_{\rm M}$), and the ellipticity of the magnetar ($\epsilon$). If the magnetic field or magnetar radius increases, the plateau luminosity is boosted, but its duration declines (see Fig~\ref{fig:flux_comparison_free_zone}, left panels). Meanwhile, as the initial period increases, the plateau luminosity decreases (see Fig~\ref{fig:flux_comparison_free_zone}, top-right panel) while the duration of the plateau grows. All these light curves mentioned above follow a decay as $L_X{\propto}t^{-2}$ after the plateau, that is, the EM losses dominate the system (see Appendix~\ref{sec:theoretical_detail}, Eqs~\ref{eq:004} and \ref{eq:006} for more details). Figure~\ref{fig:flux_comparison_free_zone}, bottom-right panel, shows light curves with different $\epsilon$ values. As $\epsilon$ increases (i.e., more significant magnetar deformation), the GWs losses become more dominant (i.e., $\epsilon{\gtrsim}$10$^{-3}$; see Appendix~\ref{sec:theoretical_detail}, Eqs~\ref{eq:005} and \ref{eq:007} for more details), and the light curve decays as $L_X{\propto}t^{-1}$ (see Eq.~\ref{eq:007}), followed by a regime wherein EM losses dominate as $L_X{\propto}t^{-2}$(see Eq.~\ref{eq:008}).

\subsection{Trapped-zone emission} \label{sec:trapped_zone}

As we explained above, in the trapped zone, part of the X-ray magnetar emission (Eq.~\ref{eq:003}) is trapped behind the ejecta material \citep[see Fig~\ref{fig:cartoon};][]{Yu2013,Sun2017}. The trapped X-ray radiation heats and accelerates the ejecta material via $pdV$ work. Under these circumstances, the optical depth ($\tau$) of the ejecta is defined as
\begin{equation}
    \tau=\kappa\left(\frac{M_{\rm ej}}{V^\prime}\right)\left(\frac{R_{\rm ej}}{\Gamma}\right),
    \label{eq:011}
\end{equation}
where $\kappa$ and $V^\prime$ are the opacity and the co-moving volume\footnote{The parameters in the co-moving frames are denoted as $Q^\prime$. Parameters without a prime ($^\prime$) are in the observer frame.}, respectively, while $\Gamma$ and $R_{\rm ej}$ are the Lorentz factor and the radius of the ejecta material, respectively. Initially, the ejecta remains extremely hot and optically thick, that is, $\tau{\gg}1$, immediately after being ejected from the vicinity of the merger. This means that the ejecta material obscures the X-ray emission of the magnetar. At the same time, the magnetar enhances the KN emission, that is, the KN is not powered by radioactive decay only but also by a fraction of the magnetar spin-down energy, rendering it brighter than a KN powered by radioactive decay alone \citep{Yu2013,Metzger2014,Metzger2019}. The KN emission boosted by the X-ray wind from magnetar spin-down is called a merger-nova \citep[a term adopted by][]{Yu2013}.
The peak luminosities of the merger-nova candidates were estimated to be above ${\gtrsim}10^{42}$~erg~s$^{-1}$, that is, one order of magnitude brighter than a standard $r$-process radioactivity-powered KN (i.e., ejecta mass and velocity of ${\sim}10^{-3}$--$10^{-4}$~$M_\odot$ and $0.3$--$0.1c$, respectively, and opacity between ${\sim}1.0$ and $10$~cm$^{2}$~$g^{-1}$). The merger-nova phenomenon is predicted to cover a wide range of peak luminosities depending on the magnetar parameters \citep{Zhang_book_2018,Yu2013}. Once the trapped magnetar wind becomes optically thin, the magnetar X-ray emission should rise quickly to a level similar to the free-zone luminosity at that point in time, and the photons can escape freely without being reprocessed by the ejecta (under the condition $\tau{<}1$).

Observationally, some KN candidates associated with SGRBs have been reported in the literature, for example, GRB~130603B \citep{Tanvir2013,Berger2013}, GRB~060614 \citep{Yang2015}, GRB~050709 \citep{Jin2016} and GRB~191019A \citep{Levan2023}, while a systematic search for KNe in GRB afterglows was done by \citet[][]{Rossi2020}; also, three compelling detections have been found: GRB~170817A \citep[e.g.,][]{Abbott2017b,Abbott2017c}, GRB~211211A \citep{Yang2022,Rastinejad2022,Troja2022b}, and recently GRB~230307A \citep{Levan2023b}. Additional systematic campaigns to search for merger-nova events have revealed three more candidates: GRBs~050724, 070714B, and 061006 \citep{Gao2017}. In the case of the GRB~170817A, very late-time observations with the \emph{Chandra} X-ray Observatory show an unabsorbed X-ray flux (luminosity) of ${\approx}1.4{\times}10^{-15}$~erg~cm$^{-2}$~s$^{-1}$ (${\approx}2.7{\times}10^{38}$~erg~s$^{-1}$), which is higher than expected from simple structured-jet model predictions, and could be explained by the energy injection from a long-lived central engine \citep[although other scenarios are not discarded completely, e.g.,][]{Troja2020,Hajela2021}. 
\citet{Sun2017} and \citet{Yu2013} have modeled the merger-nova complex system by describing the dynamical evolution of the ejecta and how the heating and cooling processes affect it, which we summarize below (for more details, see Appendix~\ref{sec:theoretical_detail}).

From the total energy of the ejecta ($E_{\rm ej}$, see Eq.~\ref{eq:012}), excluding the rest-mass energy, and considering the luminosities which boosted or lost energy to the ejecta material (see Eq.~\ref{eq:013}), such as the luminosity of the magnetar ($L_{\rm sd}$), the radioactive decay of the $r$-process material ($L_{\rm ra}$), and the bolometric luminosity of the heated electrons ($L_e$) in the system, we can derive the evolution of the Lorentz factor ($\Gamma$) of the ejecta as a function of the observer time as
\begin{equation}
    \frac{d\Gamma}{dt}=\frac{L_{\rm sd}+L_{\rm ra}-L_e-\Gamma\mathcal{D}(dE_{\rm int}^\prime/dt^\prime)}{M_{\rm ej}c^2+E_{\rm int}^\prime},
    \label{eq:014}
\end{equation}
where $E'_{\rm int}$ is the internal energy in the co-moving frame, and $\mathcal{D}{=}1/[\Gamma(1-\beta\cos\theta)]$ is the Doppler factor with $\beta{=}\sqrt{1-\Gamma^{-2}}$, $\theta{=}0^\circ$ for an on-beam observer.
The internal energy evolution should include heating from the magnetar emission and radioactivity and cooling via the electron radiation and the adiabatic work $pdV$ \citep{Kasen2010}. The evolution in the co-moving frame of the internal energy can be expressed as
\begin{equation}
    \frac{dE_{\rm int}^\prime}{dt^\prime}=\xi L_{\rm sd}^\prime+L_{\rm ra}^\prime-L_e^\prime-P^\prime\frac{dV^\prime}{dt^\prime},
    \label{eq:015}
\end{equation}
where the efficiency $\xi$ is defined as the fraction of the magnetar spin-down luminosity used to heat the ejecta, and $P'$ is the pressure (dominated by radiation) in the co-moving frame (see Eq.~\ref{eq:017}). The evolution of the internal energy also depends on the co-moving volume evolution (see Eq.~\ref{eq:018}). The definitions of $L_{\rm ra}$ and $L_e$ are provided in Appendix~\ref{sec:theoretical_detail} (see Eqs.~\ref{eq:016} and \ref{eq:020}, respectively).

The spectrum of the merger nova should approximately resemble a blackbody under the co-moving temperature, $T^\prime$,
\begin{equation}
    k_bT^\prime=\left\{
    \begin{matrix}
    k_b\left(\frac{E_{\rm int}^\prime}{aV^\prime\tau}\right)^{1/4} & {\rm for}\ \tau{>}1 \\
     & \\
    k_b\left(\frac{E_{\rm int}^\prime}{aV^\prime}\right)^{1/4} & {\rm for}\ \tau{\lesssim}1 \\
    \end{matrix}
    \right. ,
    \label{eq:021}
\end{equation}
where $k_b$ and $a$ are the Boltzmann and the blackbody radiation constants, respectively.

\begin{figure*}
    \centering
    \hspace{-0.5 cm}
    \includegraphics[scale=0.75]{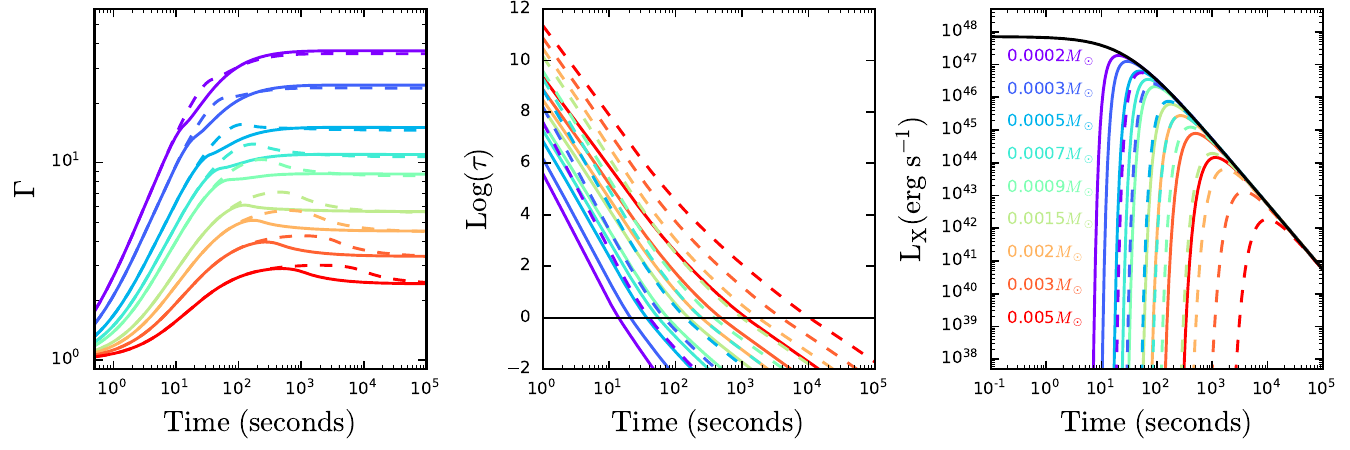}
    \vspace{-0.2 cm}
    \caption{Theoretical results of the trapped-zone emission. 
    \emph{Left panel:} evolution of Lorentz factor ($\Gamma$). 
    \emph{Middle panel:} logarithmic optical depth ($\tau$). The black horizontal line represents the optical depth limit at $\tau{=}$1.
    \emph{Right panel:} X-ray luminosity ($L_{\rm X}$) of the magnetar model (Eq.~\ref{eq:023}, first term). The solid black line shows the free zone spin-down luminosity ($\eta L_{\rm sd}$).
    The color code shows the different ejecta masses ($M_{\rm ej}{=}$2$\times$10$^{-4}$--5$\times$10$^{-3}$~$M_\odot$), while the solid and dashed lines represent opacities of 1.0 and 10.0~cm$^2$~g$^{-1}$, respectively, expected for the blue and red kilonova \citep{Tanaka2017,Metzger2019,Tanaka2020}. 
    The initial conditional assumed are: the initial velocity and internal ejecta energy of the ejecta are $\beta_i{=}$0.1, and $E_{\rm int,i}{\simeq}E_k{=}0.5M_{\rm ej}\beta_i^2c^2$, respectively. The magnetar parameters are: $B_p{=}$5$\times$10$^{15}$~G, $P_i{=}$10$^{-3}$~s, $I{=}$10$^{45}$~g~cm$^{2}$, $\epsilon{=}$10$^{-5}$, and $\xi{=}$0.3.}
    \label{fig:flux_comparison_magnetar}
\end{figure*}

For blackbody emission including the co-moving temperature $T^\prime$, the spectral luminosity at a particular frequency, $\nu$, is given by
\begin{equation}
    (\nu L_\nu)_{\rm bb}=\frac{8\pi^2\mathcal{D}^2R_{\rm ej}^2}{h^3c^2}\frac{(h\nu/\mathcal{D})^4}{\exp{(h\nu/\mathcal{D}k_bT^\prime)}-1},
    \label{eq:022}
\end{equation}
where  $h$ is Planck’s constant. 
For the trapped-zone case, the observed X-ray luminosity can be written as
\begin{equation}
    L_{\rm X,trapped}=e^{-\tau}\frac{\eta B_p^2R_{\rm M}^6\Omega(t)^4}{6c^3}+(\nu_X L_{\nu,X})_{\rm bb},
    \label{eq:023}
\end{equation}
where the first term comes from the dissipating wind, which is negligible when $\tau{\gg}$1, while the second term is the Wien tail emission of the blackbody merger-nova photosphere calculated from Eq.~\ref{eq:022} at X-ray frequencies, $\nu_X$.

Meanwhile, the specific flux of the merger nova, in the observer frame, is
\begin{equation}
    F_\nu=\frac{1}{4\pi d_L^2(1+z)}\frac{8\pi^2\mathcal{D}^2R_{\rm ej}^2}{h^3c^2\nu(1+z)}\frac{[h\nu(1+z)/\mathcal{D}]^4}{\exp{[h\nu(1+z)/\mathcal{D}k_bT^\prime]}-1},
    \label{eq:024}
\end{equation}
where $d_L$ is the luminosity distance, $z$ is the redshift, and $\nu(1+z)$ is the frequency in the source frame. It is clear that the presence of ejecta material should obscure part of the magnetar wind emission, where the shape of the X-ray light curve should depend on $t_\tau$. Nevertheless, events exhibiting a plateau at early times could still be interpreted from a trapped-zone emission perspective. 

We analyze different cases for comparison with the free-zone emission, where the plateau has a duration of $T_X$ and the trapped-zone emission has an obscuration time $t_\tau$:
\begin{enumerate}
    \item Cases with $t_\tau{\ll}T_X$ could be related to sources with a very low ejecta mass (i.e., the line of sight passes through the trapped zone where the ejecta masses are still low), such that the resulting light curves are similar to the free-zone case.

    \item For $t_\tau{\gg}T_X$ (i.e., the line of sight passes deep into the trapped zone), the resulting light curve (and spectra) will strongly differ from the free zone case, and we expect to observe just (part of) the decay phase. 

    \item If $t_\tau{\sim}T_X$, we may observe just a small final portion of the plateau emission, followed by the decay phase.
\end{enumerate}

If the X-ray emission is viewed from the free zone, the X-ray plateau approximately begins with the BNS merger coalescence. 
Then, the multi-wavelength emission of the merger-nova at $t{<}t_{X,e}$, where $t_{X,e}$ is the emerging time of the X-rays, is not important because the magnetar emission and the merger nova should be approximately simultaneous. On the other hand, if the X-ray emission is observed from the trapped zone, the merger time should be $t_{X,e}{-}t_\tau$, that is, the multi-wavelength observations of the merger nova at $t{<}t_{X,e}$ become relevant \citep{Ai2021}.

Numerical solutions to the above equations for the free- and trapped-zone emissions are shown in Fig.~\ref{fig:flux_comparison_magnetar} for a set of ejecta masses from 2${\times}$10$^{-4}$ to 5$\times$10$^{-3}$~$M_\odot$ (color coded), and two different opacities $\kappa{=}$1.0 (solid lines) and 10~cm$^2$~g$^{-1}$ (dashed lines) under some fixed initial conditions (see the caption of Fig.~\ref{fig:flux_comparison_magnetar}).
Specifically, Fig.~\ref{fig:flux_comparison_magnetar}, left panel, shows the dynamical evolution of the ejecta material (i.e., the Lorentz factor) for different masses.
The Lorentz factor increases substantially while $t{\lesssim}t_\tau$, beyond which point the rate slows and flattens near some maximum, with a bit of an overshoot observed for higher ejecta masses. The slope of the rise in $\Gamma$ and the time to reach maximum are both strong functions of the ejecta mass (with lower ejecta masses associated with more highly relativistic ejecta). These trends are expected because the ejecta becomes optically thin, and the magnetar emission does not influence the dynamical evolution at $t{>}t_\tau$. 
Figure~\ref{fig:flux_comparison_magnetar}, middle panel, depicts the optical depth evolution, which is high at early times and shows a smooth, strong decline with time as the ejecta expands, eventually passing from the optically thick to thin regimes around $t{\approx}t_\tau$. This transition is an important factor for observing the magnetar emission. In Fig.~\ref{fig:flux_comparison_magnetar} right panel, we see that the X-ray luminosity of the magnetar exhibits an extremely steep rise at early times, reaching a peak luminosity at $t{\sim}t_\tau$, after which the ejecta is optically thin, allowing the magnetar X-ray photons to escape freely.
In Sects.~\ref{sec:opacity} and \ref{sec:mergernova}, we discuss the role of opacity in the X-ray light curve and the merger-nova emission, respectively.

\begin{figure*}
    \centering
    \hspace{-0.5 cm}
    \includegraphics[scale=0.85]{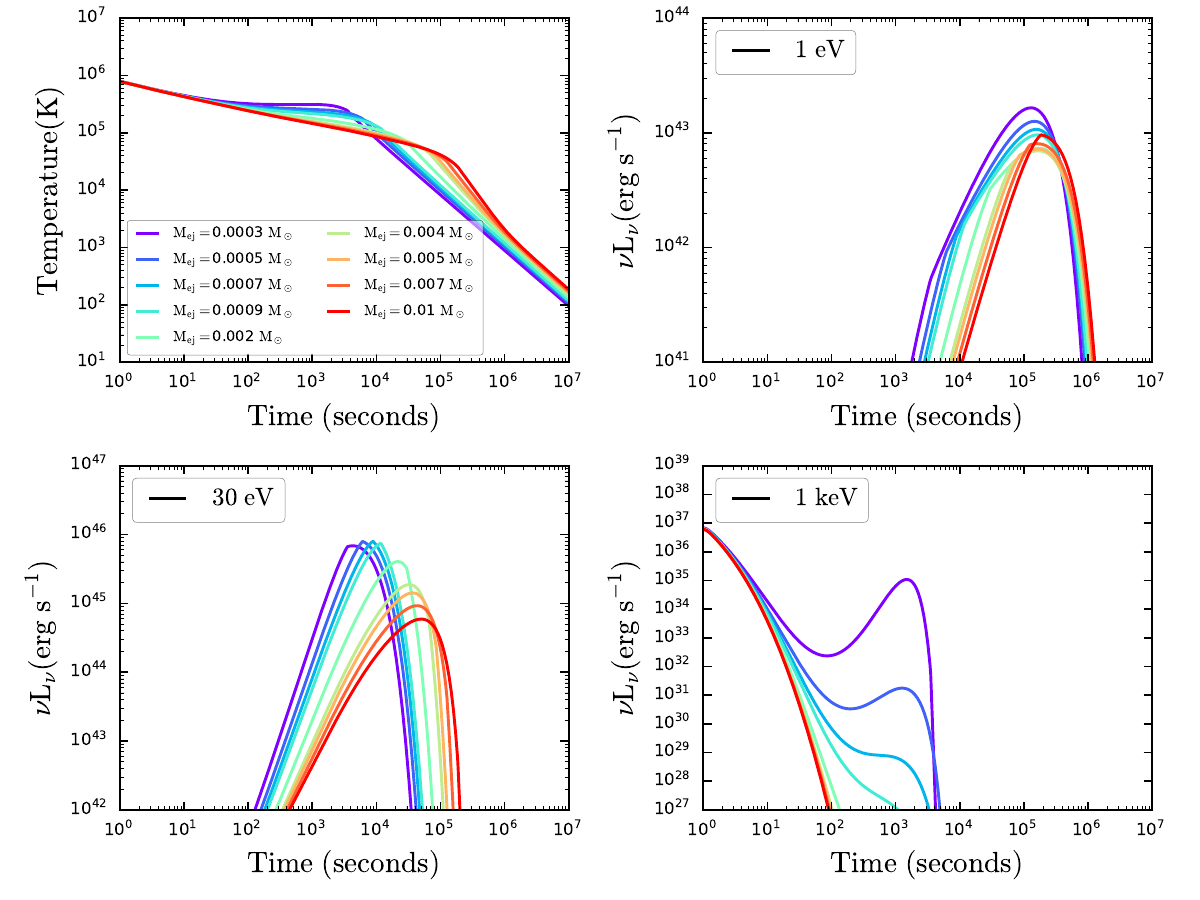}
    \vspace{-0.5 cm}
    \caption{Evolution of the thermal component and light curves of the merger-nova emission (Eq.~\ref{eq:022}) for different ejecta masses, color-coded from 10$^{-4}$ to 5$\times$10$^{-3}$~$M_\odot$. The top-left panel is the temperature evolution in the observer frame, while the other three panels represent the merger-nova emission at different frequencies (i.e, $h\nu$: 1~keV, 30~eV, and 1~eV). The model fixed parameters are $\xi{=}0.3$, $\eta=10^{-3}$, $\kappa{=}1.0$~cm$^{2}$~g$^{-1}$, $B_p{=}3\times10^{15}$~G, and $P_i=5$~ms.}
    \label{fig:flux_margernova}
\end{figure*}

\subsubsection{Opacity} \label{sec:opacity}

There are several uncertainties associated with the expected opacity adopted in theoretical models and from observations of AT2017gfo \citep[e.g.,][]{Tanaka2017,Tanaka2020,Banerjee2020,Sarin2021}. For instance, \citet{Banerjee2020}, from simulations, argued that the Planck mean opacity for lanthanide-free ejecta (so-called blue KN) could be in the range $\kappa{\approx}$1--0.5~cm$^2$~g$^{-1}$ at $t{\sim}$0.1~days, and could increase up to $\kappa{\approx}$10--5~cm$^2$~g$^{-1}$ at $t{\sim}$1~day. 
In other words, the opacity of the ejecta changes with time. Observationally, lanthanide-free ejecta material ($\kappa{\approx}$5--3~cm$^2$~g$^{-1}$) is needed to explain the early blue KN emission from AT2017gfo, while lanthanide-rich material ($\kappa{\approx}$30--20~cm$^2$~g$^{-1}$) is necessary to reproduce the more long-lasting NIR emission of AT2017gfo \citep[e.g.,][]{Tanaka2020}. Thus, currently, it is accepted that AT2017gfo had at least two components: a red and a blue KN \citep{Arcavi2017,Cowperthwaite2017,Chornock2017,Drout2017,Evans2017,Nicholl2017,Smartt2017,Kasen2017,Kilpatrick2017,Shappee2017,Tanvir2017,Gao2017}. Nevertheless, some authors \citep[such as][from semi-analytical models]{Villar2017} suggested that the data can be well modeled by including a three-component KN with an intermediate (``purple'') opacity component ($\kappa{=}3$~cm$^2$~g$^{-1}$).

Figure~\ref{fig:flux_comparison_magnetar} explores the influence of the opacity on the dynamical evolution of the ejecta and the X-ray light curves considering opacities of $\kappa{=}$1.0 (solid lines) and 10.0~cm$^2$~g$^{-1}$ (dashed lines). We see that an increase in the opacity leads to a more extended overshoot in the Lorentz factor (see Fig.~\ref{fig:flux_comparison_magnetar}, left panel), which is more noticeable for higher ejecta masses and longer timescales for the ejecta to become optically thin (middle panel) and reach a lower X-ray peak luminosity (right panel).

\subsubsection{Merger-nova emission} \label{sec:mergernova}

Even if only a modest fraction of the rotational energy liberated in the spin down (see Eq.~\ref{eq:001}) is converted to EM radiation in the hours to years after the merger, this would substantially enhance the EM luminosity of the merger counterpart \citep[][and references therein]{Metzger2019}.
A merger-nova should emit quasi-blackbody-like emission at a specific temperature (given by Eq.~\ref{eq:021}), generating emission (given by Eqs.~\ref{eq:022} and \ref{eq:024}) across the EM spectrum. The light curves at X-ray, UV, and optical wavelengths (i.e., at 1~keV, 30~eV, and 1~eV energies, respectively) and the temperature evolution of the merger-nova are shown in Fig.~\ref{fig:flux_margernova}. The color code represents different ejecta masses (from 3${\times}$10$^{-4}$ to 1${\times}10^{-2}$~M$_\odot$), assuming typical parameters (see caption in Fig.~\ref{fig:flux_margernova}). 

\begin{figure*}
    \centering
    \includegraphics[scale=0.75]{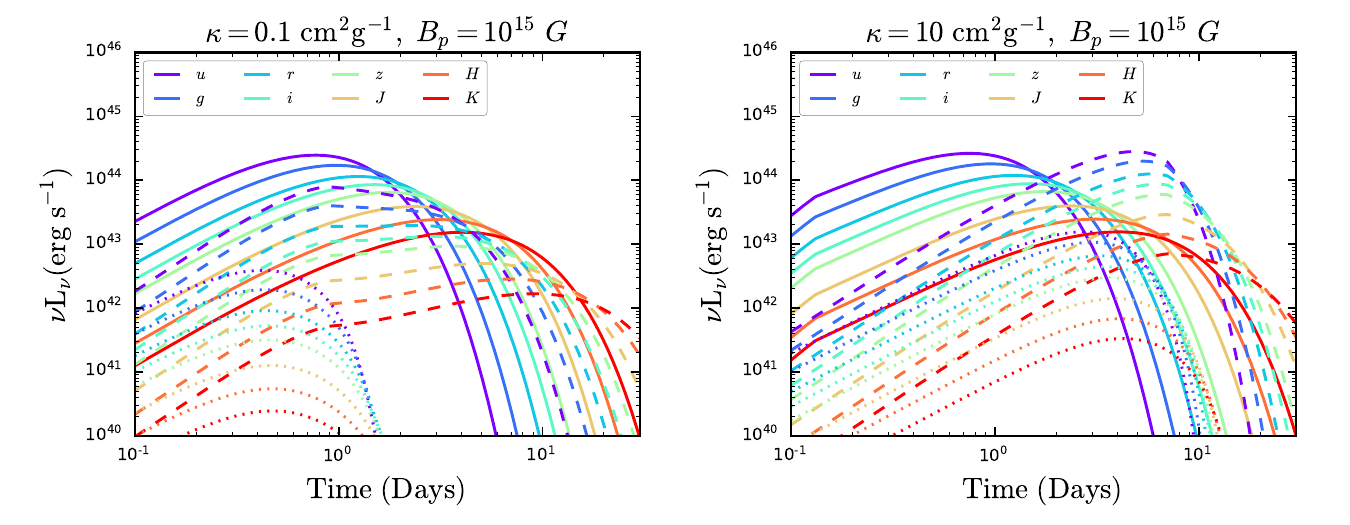}
    \includegraphics[scale=0.75]{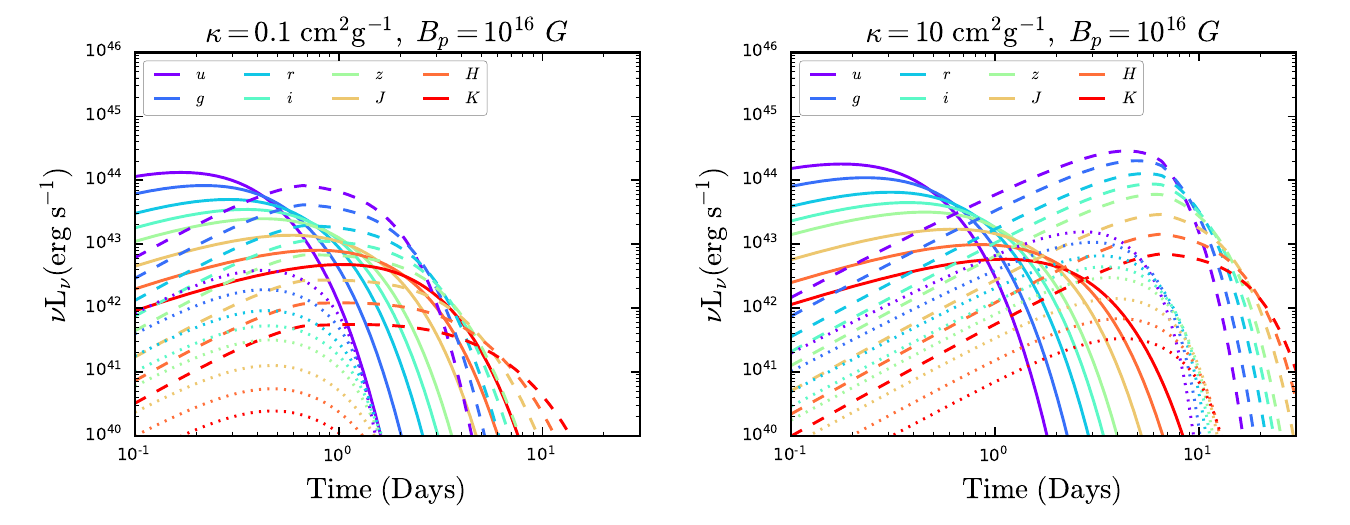}
    \vspace{-0.3 cm}
    \caption{Theoretical light curves of the merger-nova emission, across several common optical/NIR filters, considering opacities of $\kappa{=}$0.1 (left panels) and 10~cm$^{2}$~g$^{-1}$ (right panels), representing the lanthanide-free and -rich ejecta cases, respectively. KN emission without any magnetar spin-down enhancement is shown for comparison (dotted lines). The model fixed parameters are $P_i{=}$5~ms, $\xi{=}0.3$, and $\eta=10^{-3}$, along with two adopted ejecta masses of $M_{\rm ej}{=}10^{-4}$ (solid lines) and $10^{-2}$~M$_\odot$ (dashed lines). The upper and lower panels adopt magnetic field strengths of $B_p{=}10^{15}$~G and $10^{16}$~G, respectively.}
    \label{fig:margernova_filters}
\end{figure*}

In general, the temperature of the merger-nova decreases with time (see Fig~\ref{fig:flux_margernova}, top-left panel), from ${\approx}10^6$ (at ${\sim}1$~s) to ${\approx}10^3$~K (at ${\sim}10^6$~s), causing the peak of the blackbody emission to shift from X-ray to NIR wavelengths over this time period. The temperature decrease is initially rather shallow up to the pivot point, which coincides with the time when the ejecta changes from optically thick to thin, followed by a faster decay. The effect of the magnetar on the enhancement of the temperature is stronger for low-ejecta masses (see Fig~\ref{fig:flux_margernova}, top-left panel).

In detail, the 1~keV X-ray emission of the merger nova dominates at early times, with a peak luminosity of $L_{\rm X,peak}{\lesssim}$10$^{37}$~erg~s$^{-1}$ at 1~s
(see Fig~\ref{fig:flux_margernova}, bottom-right panel). The relatively low (${<}10^{6}$~K) and decreasing temperature as the ejecta cools leads to a strong temporal decline as the Wien tail of the blackbody emission passes from the X-ray to UV to optical bands.
The blackbody merger-nova luminosity is highest in the UV band (30~eV), reaching peak luminosities of ${\approx}$10$^{45}$~erg~s$^{-1}$ at ${\approx}10^{4}$--$10^{5}$~s (see Fig~\ref{fig:flux_margernova}, bottom-left panel).
The optical light curves (1~eV) reach peak luminosities of ${\approx}$10$^{43}$--$10^{44}$~erg~s$^{-1}$ on timescales of 4--5 days (see Fig~\ref{fig:flux_margernova}, top-right panel).

Figure~\ref{fig:margernova_filters} shows the optical/NIR merger-nova light curves for the $u$- to $K$-bands (derived using Eq.~\ref{eq:022}), considering a magnetic field strength and initial rotational period of $B_p{=}10^{15}$~G and $P_i{=}$5~ms, respectively, and ejecta masses of $M_{\rm ej}{=}10^{-4}$ (solid lines) and $M_{\rm ej}{=}10^{-2}$~M$_\odot$ (dashed lines). KN light curves with no magnetar spin-down enhancement are shown for comparison (dotted lines). Both low and high values for the ejecta opacity of $\kappa{=}0.1$ (left panels) and $\kappa{=}20$~cm$^2$~g$^{-1}$ (right panels), representing lanthanide-free and lanthanide-rich ejecta scenarios, respectively, are shown.

We see that the emission peaks earlier in the bluer filters, with little dependence on ejecta mass (see Fig.~\ref{fig:margernova_filters}), similar to AT2017gfo \citep[e.g.,][]{Kasen2017,Villar2017}.
For the low-opacity (lanthanide-free) ejecta case, the time to peak only changes by a factor of $\sim$2--3 between the low ($M_{\rm ej}{\sim}10^{-4}$~M$_\odot$) to high ($M_{\rm ej}{\sim}10^{-2}$~M$_\odot$) ejecta masses, with bluer filters showing slightly smaller shifts. Meanwhile, the luminosities decrease in all filters by a factor of ${\sim}5$ for a 100-fold increase in ejecta mass. 
By contrast, for the higher opacity (lanthanide-rich) case, the time to peak is extended by a factor of $\sim$10 for the bluest and $\sim$3 for the reddest filters for the higher ejecta mass scenario (see Fig.~\ref{fig:margernova_filters}, right panels), while the peak luminosities do not change considerably.

\begin{table}
    \centering
    \caption{Parameters and uniform priors distributions considered for the free and trapped magnetar models.}
    \scalebox{1.0}{
    \begin{tabular}{lll}
    \hline\hline
    Parameter & Model zone & Prior \\ \hline
    (1) & (2) & (3) \\ \hline
    $\log(B_P/{\rm G})$ & Free/Trapped & $\mathcal{U}{\sim}\left[14;17\right]$\\
    $\log(P_i/{\rm s})$ & Free/Trapped & $\mathcal{U}{\sim}\left[-4;-2\right]$\\
    $\log(M_{\rm ej}/M_\odot)$ & Trapped & $\mathcal{U}{\sim}\left[-4;-2\right]$\\
    $\log(\kappa/{\rm cm}^{2}\ {\rm g}^{-1})$ & Trapped & $\mathcal{U}{\sim}\left[-2;2\right]$\\
    $\log(T_0/{\rm s})$ & Trapped & $\mathcal{U}{\sim}\left[-1;2.5\right]$\\
    $\log(\xi)$ & Trapped & $\mathcal{U}{\sim}\left[-4;0\right]$\\
    \hline
    \end{tabular}
    }
    \label{tab:fitting_para_stat}
\end{table}

The lower panels of Fig.~\ref{fig:margernova_filters} consider a higher magnetic field of $B_p{=}10^{16}$~G. An enhanced magnetic field implies a lower EM characteristic spin-down timescale (see Eq.~\ref{eq:004}), and an increased luminosity (see Eq.~\ref{eq:006}), as the magnetar should lose rotational energy more quickly, and its contribution to the ejecta emission is enhanced in early epochs, although it is brief. This is shown in Fig.~\ref{fig:margernova_filters}, where the merger-nova emission is enhanced at early epochs in blue and red filters. In contrast, the magnetar does not give energy to the ejecta at later epochs, and the emission becomes fainter and decays faster.

Finally, according to \citet{Metzger2019}, the contribution of the magnetar emission could have three different effects on the light curves of normal KNe, which are visible in Fig.~\ref{fig:margernova_filters}: $i)$ an increase in the peak luminosity by a few orders of magnitude due to additional heating from the magnetar spin-down; $ii)$ more rapid evolution, that is, an earlier peak time because the greater ejecta velocity due to the kinetic energy added to the ejecta by the spin-down energy released during the optically thick phase; and $iii)$ emission shifts toward bluer energy bands because the high luminosity of the transient increases its effective temperature. These effects only consider the optimistic case where the magnetar resists an immediate collapse to form a BH or an essential fraction of its rotational energy does not escape as gamma-rays \citep[e.g.,][]{DallOsso2009,Corsi2009}.

\section{Fast X-ray transient sample} \label{sec:data}

We consider now the sample of 22 FXT candidates reported in Papers I and II, 
which can be crudely separated according to their distances: $i)$ 5 nearby FXTs located at ${\lesssim}$100~Mpc, and $ii)$ 17 distant FXTs located at ${\gtrsim}$100~Mpc. The five nearby FXTs have peak luminosities of $L_{\rm X,peak}{\lesssim}10^{40}$~erg~s$^{-1}$. Among the 17 distant FXTs, 8 have been associated with faint spatially resolved galaxies (with spectroscopic or photometric redshifts in the range of ${\sim}$0.7--2.2), implying peak luminosities of $L_{\rm X,peak}{\gtrsim}10^{44}$~erg~s$^{-1}$, and a remaining 9 have no clear host galaxies identified so far.

In this work, we do not consider the nearby sample because their low luminosity peak ($L_{\rm peak}{\lesssim}10^{40}$~erg~s$^{-1}$) does not match with the BNS emission. In particular, we explore for a subset of the distant FXTs whether they can be described by the spin-down power of a massive magnetar formed in the immediate aftermath of a BNS merger, as modeled by the equations in Sect.~\ref{sec:theoretical}.
Notably, the light curves in the free and trapped zones (see Figs.~\ref{fig:flux_comparison_free_zone} and \ref{fig:flux_comparison_magnetar}, respectively) are different, such that some distant FXTs may remain more consistent with free zone viewing-angle models, whereas others may be more consistent with trapped zone viewing-angle models. Among the 17 distant FXTs identified in \emph{Chandra} data, six (FXTs~7, 10, 12, 16, 19, and 22) show early plateaus in their light curves (potential free-zone cases) while three sources (FXTs~14, 15, and 20) show clear rising and falling behavior (potential trapped-zone cases).
The other eight \emph{Chandra} FXTs do not follow the expected light curves\footnote{We discarded events that do not follow the expected light curve decays: both XRT~000519 and 110103 have a substructure in their main peaks \citep{Jonker2013,Glennie2015,Quirola2022} which does not keep an explanation under the magnetar model; in contrast, although the others have a power-law decay, their slopes are not consistent with the expected by the magnetar model (see Sects.~\ref{sec:free_zone} and \ref{sec:trapped_zone})
.} of this interpretation, which suggests a different origin, for instance, IMBH TDEs. 

Before exploring the parameter space that explains the FXT light curves, we briefly review their current distances, which are listed in Table~\ref{tab:fitting_para} (column 2). FXT~16 has a firm spectroscopic host redshift of $z{=}0.738$ \citep{Xue2019}, while FXTs~14 and 19 have constrained photometric host redshifts of $2.23_{-1.84}^{+0.98}$ and $1.44{\pm}0.08$, respectively \citep{Bauer2017,Lin2022,Quirola2023}. FXT~22 has three possible host galaxy associations, denoted as cX, cNE and cW \citep[at 0\farcs6, 4\farcs2 and 3\farcs6 from the transient position, respectively;][]{Eappachen2023}; unfortunately, cX (which is formally inside the X-ray uncertainty region) has only a single detection at $g{=}25.9{\pm}0.1$~AB~mag, precluding any photometric and spectroscopic redshift determination, while, cNE and cW have spectroscopic and photometric redshifts of $z_{\rm cNE}{=}1.5082{\pm}0.0001$ \citep{Andreoni2021a,Jonker2021} and $z_{\rm cW}{=}1.04^{+0.22}_{-0.14}$ \citep{Eappachen2023}, respectively.
 For FXTs for which no clear host galaxy has been identified so far, we adopt two nominal redshifts of 0.5 and 1.0 (see Papers~I and II for details).

\begin{table*}
    \centering
    \caption{Parameters obtained from the fitting process of the magnetar model.}
    \hspace{-0.8cm}
    \scalebox{0.75}{
    \begin{tabular}{llllllllllll}
    \hline\hline
    FXT & ID & $z$ & Free/Trapped & $\log(\kappa/{\rm cm}^{2}\ {\rm g}^{-1})$ & $\log(M_{\rm ej}/M_\odot)$ & $\log(T_0/{\rm s})$ & $\log(\xi)$ & $\log(B_P/{\rm G})$ & $\log(P_i/{\rm s})$ & $\chi^2$/dof & BIC/AIC \\ 
    (1) & (2) & (3) & (4) & (5) & (6) & (7) & (8) & (9) & (10) & (11) & (12) \\ \hline
    \multicolumn{12}{c}{FXTs with known redshift} \\ \hline 
    14$^{\dagger}$ & XRT~141001 & $2.23$ & Trapped &  $-1.0_{-0.47}^{+1.39}$ & $-3.67_{-0.79}^{+0.38}$ & $1.33_{-1.85}^{+0.21}$ & $-2.86_{-0.16}^{+0.24}$ & $15.73_{-0.05}^{+0.14}$ & $-2.84_{-0.15}^{+0.10}$  & $11.14/23$  & $-7.55/-15.75$ \\   
    16 & XRT~150322 & $0.738$ & Free & -- & -- & -- & -- & $15.28{\pm}0.04$  & $-2.57_{-0.02}^{+0.03}$  & $41.21/28$ & $16.33/13.53$  \\    
    19 & XRT~170901 & $1.44$ & Free & -- & -- & -- & -- & $14.84_{-0.10}^{+0.07}$ & $-2.98_{-0.03}^{+0.04}$  &  $43.06/39$ & $9.43/6.01$  \\
    22$^{\dagger\dagger}$ & XRT~210423 & $1.5105$ (cNE) & Free & -- & -- & -- & -- & $14.86_{-0.12}^{+0.09}$  &  $-2.93_{-0.03}^{+0.05}$ & $29.03/27$  & $6.77/4.03$  \\    
     & & $1.04$ (cW) & Free & -- & -- & -- & -- & $14.92_{-0.12}^{+0.07}$  & $-2.82_{-0.04}^{+0.03}$  & $29.03/27$  &  $6.76/4.03$ \\
    \hline
    \multicolumn{12}{c}{FXTs with nominal redshift} \\ \hline
    7 & XRT~030511 & $0.5$ & Free & -- & -- & -- & -- & $15.19{\pm}0.05$  & $-2.58{\pm}0.02$  & $62.97/40$  & $24.49/21.01$  \\
     & &  $1.0$ & Free & -- & -- & -- & -- & $15.01_{-0.05}^{+0.04}$  & $-2.83{\pm}0.02$  & $62.95/40$  &  $24.47/21.0$ \\
    10 & XRT~100831 & $0.5$ & Free & -- & -- & -- & -- & $15.64_{-0.36}^{+0.15}$ & $-2.10_{-0.10}^{+0.06}$  & $4.37/7$ & $-2.1/-2.5$  \\
     & & $1.0$ & Free & -- & -- & -- & -- & $15.46_{-0.22}^{+0.15}$ &  $-2.35_{-0.08}^{+0.09}$ & $4.36/7$  & $-2.12/-2.52$  \\
    12 & XRT~110919 & $0.5$ & Free & -- & -- & -- & -- & $15.53_{-0.07}^{+0.08}$ & $-2.27{\pm}0.04$ & $14.71/21$  &  $-4.0/-6.28$ \\
     & & $1.0$ & Free & -- & -- & -- & -- & $15.35{\pm}0.07$  & $-2.52_{-0.03}^{+0.04}$  & $14.72/21$  & $-4.0/-6.27$ \\
    15 & XRT~140507 & $0.5$ & Trapped & $-0.66$(fix) & $-2.50_{-2.09}^{+0.74}$ & ${<}2.49$ & $-0.28$(fix) & $15.77_{-0.11}^{+0.15}$ & $-3.0_{-0.46}^{+0.72}$ & $2.36/2$ & $1.58/2.41$ \\
     & & $1.0$ & Trapped &  $-1.0$(fix) & $-2.50_{-1.58}^{+0.78}$ & $2.36_{-2.27}^{+0.05}$ & $-1.80$(fix)  & $15.60_{-0.09}^{+1.09}$ & $-3.0_{-0.56}^{+0.57}$  & $2.22/2$ & $1.21/2.04$  \\
    20 & XRT~191127 & $0.5$ & Trapped & ${<}1.6$ & $-3.53_{-0.29}^{+1.54}$  &  $0.83_{-1.52}^{+0.37}$  & ${<}0.0$ & $16.16_{-1.24}^{+0.12}$  &  $-3.15_{-0.23}^{+0.66}$ & $5.62/3$  & $8.95/7.77$  \\
     & & $1.0$ & Trapped & $-0.97_{-0.70}^{+1.76}$ & ${<}-1.62$  &  $0.84_{-0.86}^{+0.97}$  & ${>}-3.60$ & $16.02_{-0.88}^{+0.08}$  &  ${>}-3.83$ & $5.19/3$  & $8.22/7.04$  \\
    \hline
    \end{tabular}
    }
    \tablefoot{\emph{Column 1:} FXT candidate number, from Papers I and II. 
    \emph{Column 2:} ID. 
    \emph{Column 3:} Redshift. 
    \emph{Column 4:} Free or trapped models considered by the magnetar model.
    \emph{Column 5:} Opacity of the ejecta material. 
    \emph{Column 6:} Mass of the ejecta material. 
    \emph{Column 7:} $T_0$ parameter.
    \emph{Column 8:} Efficiency factor $\xi$ is the fraction of the magnetar spin-down luminosity transferred to heat the ejecta material.
    \emph{Column 9:} Dipole component of the magnetic field at the poles.
    \emph{Column 10:} Initial period of the magnetar.
    \emph{Columns 11 and 12:} Chi-square/degree-of-freedom and Bayesian/Aikake information criteria (BIC/AIC), respectively.\\
    $^{\dagger}$ FXTs with known photometric redshifts \citep{Bauer2017}.\\
    $^{\dagger\dagger}$ Spectroscopic and photometric redshift for two of the three candidate host galaxies of XRT~210423 \citep{Eappachen2023}.
    }
    \label{tab:fitting_para}
\end{table*}

\section{Fitting data} \label{sec:fitting}

We fit the observed light curves using the free and trapped models to derive the magnetar and ejecta parameters. For the fitting process, we assume typical values, based on simulations and previous observational constraints, of $R_{\rm M}{=}1.2{\times}10^6$~cm, and $\eta{=}$10$^{-3}$ \citep[e.g.,][reported values of $\eta{\sim}10^{-3}$ for the magnetar emission of CDF-S XT1 and XT2\footnote{In Appendix~\ref{sec:redshift_efficiency}, we discuss the role of the efficiency in deriving the magnetar parameters.}]{Sun2019}, and the initial velocity, radius and internal energy of the ejecta of $\beta_i{=}0.1$ (i.e., initial Lorentz factor of $\Gamma_i{=}$1.005), $R{_{\rm ej}}_{,i}{=}2.4{\times}10^7$~cm \citep[e.g.,][]{Radice2018b,Radice2018a}, and $E_{\rm int,i}{\simeq}E_k{=}0.5M_{\rm ej} (\beta_i c)^2$, respectively, where $E_k$ is the kinetic energy. The assumed parameters do not affect the results of the fitting. For instance, a change in the initial velocity and radius of the ejecta of $\beta_i{=}0.3$ and $R{_{\rm ej}}_{,i}{=}4.8{\times}10^7$~cm (i.e., an increase by a factor of ${\approx}3-2$), respectively, means a switch in the peak luminosity of ${\lesssim}1$\%.
Spherical symmetry amongst the ejecta material is assumed, as we expect the ejecta to expand radially over many orders of magnitude from ${\sim}10^6$ to ${\sim}10^{15}$~cm \citep{Metzger2019}. We perform the light-curve fitting with the least squares method, using the \texttt{Python} package \textsc{lmfit}, while to determine the uncertainties of the parameters, we use a Bayesian inference from a  Monte Carlo method \citep{Foreman_Mackey2013,Sharma2017} and the \textsc{MultiNest} nested sampling algorithm \citep{Feroz2008,Feroz2009}.
We assume uniform logarithmic prior distributions for each free parameter (see Table~\ref{tab:fitting_para_stat} for more details).

The two magnetar parameters $B_{\rm P}$ and $P_i$ are free parameters in both the free-zone and trapped-zone models, while three ejecta parameters ($M_{\rm ej}$, $\kappa$ and $\xi$) in the trapped zone model are also free in the fitting process. However, due to the low number of bins in the light curve of FXT~15, we freeze some parameters during the fitting process to retain a positive number of degrees of freedom (dof).

Notably, in the trapped-zone scenario, the peak emission depends strongly on both the $M_{\rm ej}$ and the time when the light curve starts (i.e., the BNS merger). As this is not necessarily known, we introduce an additional parameter called $T_0$, which attempts to correct the start point. In the free and trapped models, we assume an ellipticity parameter of $\epsilon{=}10^{-5}$ as a consequence of the fact that the light curve decays are not consistent with $F_X{\propto}t^{-1}$ (see Eq.~\ref{eq:007}). 

The best-fitting parameters are shown in Table~\ref{tab:fitting_para}, including 5,000 simulations followed to convergence to determine errors. The uncertainties in the parameters represent the 16\% and 84\% percentiles of the Markov Chain Monte Carlo (MCMC) distributions. The $\chi^2$, dof, Bayesian Information Criterion (BIC), and Akaike Information Criteria (AIC) are derived from the best-fitted parameters under the least squares method.
In sections~\ref{sec:free_FXTs} and \ref{sec:trapped_FXTs}, we describe the results obtained for the free- and trapped-zone samples, respectively.

\begin{figure*}
    \centering
    \includegraphics[scale=0.85]{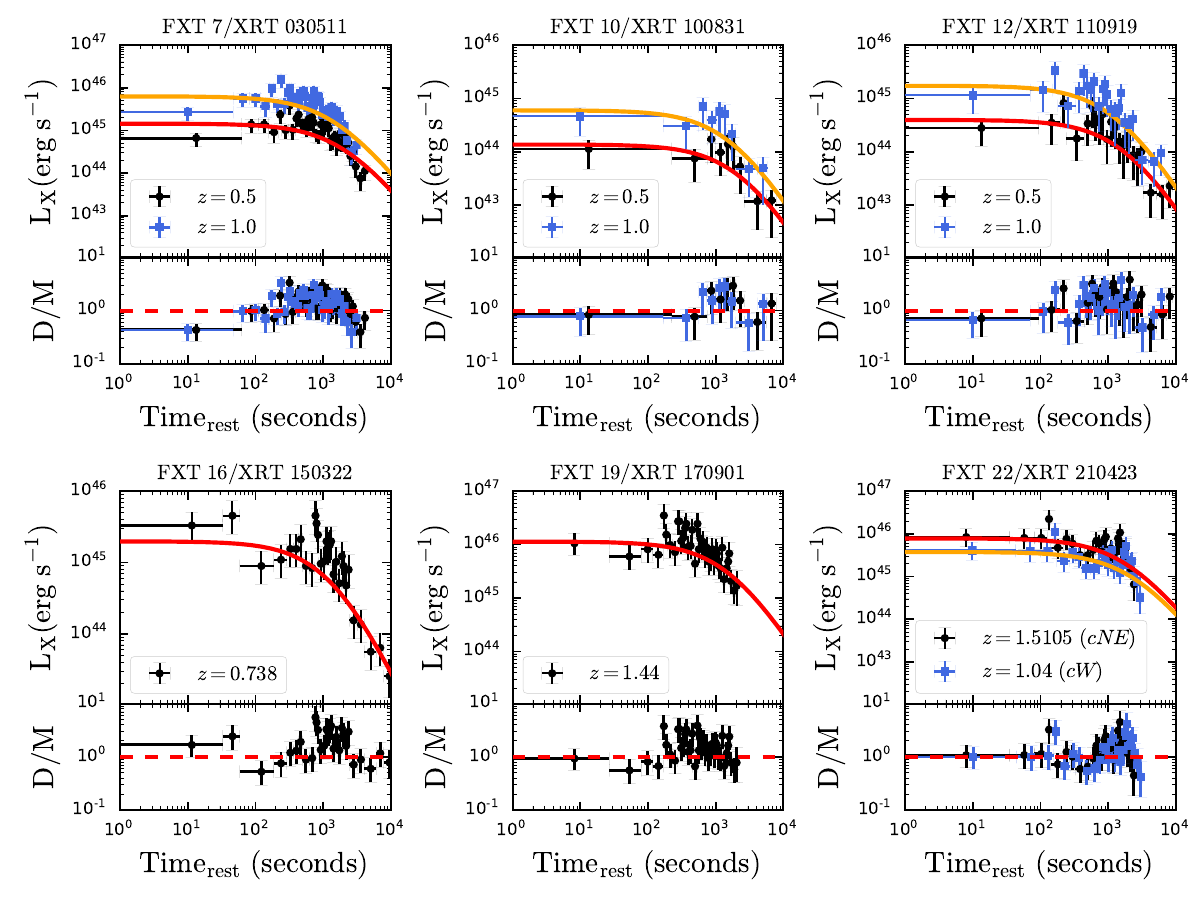}
    \vspace{-0.3cm}
    \caption{Free-zone fitting (solving the system of Eqs.~\ref{eq:010} and \ref{eq:002}) in the source rest-frame for FXTs with known and nominal redshifts (see Table~\ref{tab:fitting_para}). Time dilation effects correct the data, and the bottom panels show the ratio between the data and the model (D/M).}
    \label{fig:fitting_free}
\end{figure*}

\subsection{Free-zone FXTs}\label{sec:free_FXTs}

The light curves of known-redshift FXTs~16, 19, and 22 (assuming either $z_{cW}{=}1.04$ or $z_{cNE}{=}1.5105$) are well-fitted with the free-zone model. The best-fit models and their residuals are shown in Fig.~\ref{fig:fitting_free}, while the posterior distributions can be found as 
Figs.~\ref{fig:Ch4_posterior_free_FXT1619} and \ref{fig:Ch4_posterior_free_FXT22}; the latter suggest a correlation between $B_{\rm P}$ and $P_i$ with a coefficient between 0.6 and 0.9.
The best-fit magnetar parameters cover a narrow region in the parameter space of $B_P{\approx}3{\times}10^{15}$--$7{\times}10^{14}$~G and $P_i{\approx}7.9$--$1.2$~ms, and are consistent with literature-derived values in the cases of FXT~16 \citep[i.e., $B_P{\sim}6{\times}10^{15}$~G and $P_i{\sim}4.4$~ms;][]{Xue2019,Xiao2019,Lu2019} and FXT~22 \citep[i.e., $B_P{\lesssim}7{\times}10^{14}$~G;][]{Ai2021}.

For the hostless FXTs~7, 10, and 12, we assume two nominal redshifts of $z_{\rm nominal}{=}$0.5 and 1.0\footnote{In Appendix~\ref{sec:redshift_efficiency}, we discuss the role of the nominal redshifts in deriving the magnetar parameters.}. These FXTs are well-fitted by the free-zone model, with best-fitting models and residuals shown in Fig.~\ref{fig:fitting_free} and posterior distributions in Figs.\ref{fig:Ch4_posterior_free_FXT7}--\ref{fig:Ch4_posterior_free_FXT12}.
For FXTs~7 and 12, \citet{Lin2022} also identified that the EM losses are the main contribution to explaining the light curve trends, discarding GW losses as the prime route to dissipating rotational energy. It is important to realize that increasing the redshift from $z{=}0.5$ to $z{=}1.0$ results in slight drops in both $B_{\rm P}$ and $P_i$ (see Table~\ref{tab:fitting_para}). Thus, if these FXTs lie at larger distances, they will likely have lower magnetar parameters, although these are presumably bounded by the NS breakup limit at ${\sim}0.96$~ms \citep{Lattimer2004,Rowlinson2013}.

\begin{figure*}
    \centering
    \hspace{-1.2cm}
    \includegraphics[scale=0.9]{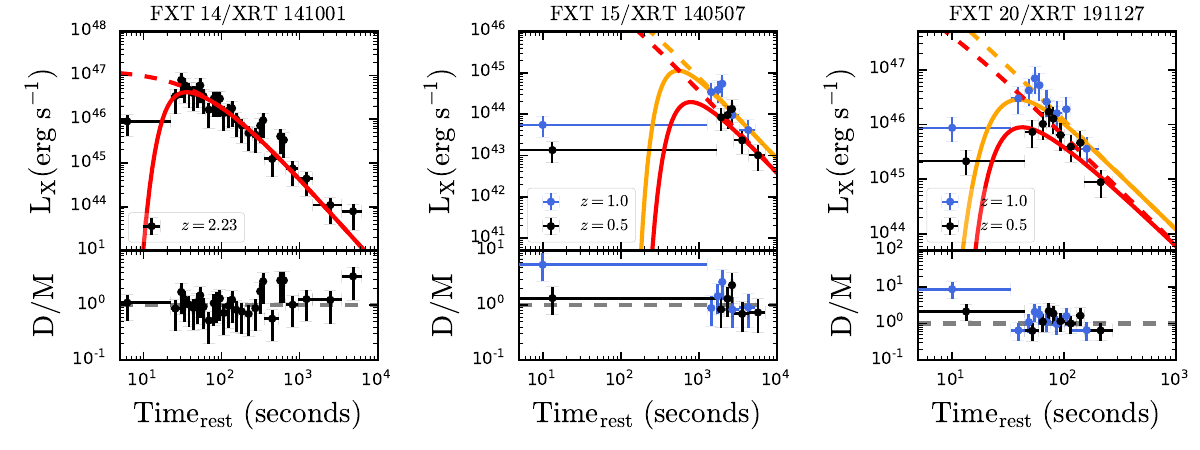}
    \vspace{-0.2cm}
    \caption{Trapped zone light-curve fitting of luminosities (solving the system of Eqs.~\ref{eq:014}--\ref{eq:023}) in the source rest frame for FXTs with known and nominal redshifts (see Table~\ref{tab:fitting_para}). The bottom panels show the ratio between the data and the model values (D/M). The dashed lines represent the unobscured magnetar emission.}
    \label{fig:fitting_trapped}
\end{figure*}

\subsection{Trapped-zone FXTs}\label{sec:trapped_FXTs}

The light curve of known-redshift FXT~14 is well-fitted with the trapped-zone model. The best fit and residuals are shown in Fig.~\ref{fig:fitting_trapped}, while the posterior distributions can be found in Fig.~\ref{fig:MCMC_trapped}. Notably, the posterior distributions of FXT~14 imply that degeneracies exist between the parameters $T_0$, $\xi$, and $\kappa$. Although the ejecta appears to be low mass ($M_{\rm ej}{\sim}2{\times}10^{-4}$~M$_\odot$) and lanthanide-free ($\kappa{\sim}0.1$~cm$^2$~g$^{-1}$), the uncertainties for the opacity are significant (${\sim}$6--0.03~cm$^2$~g$^{-1}$), spanning the values adopted or reported in the literature \citep[i.e., $\kappa{\approx}1$~cm$^2$~g$^{-1}$;][]{Sun2019}. The derived magnetar parameters are also consistent with the values obtained in previous works \citep[e.g., $B_P{\approx}10^{16}$~G and $P_i{\approx}1.2$~ms;][]{Sun2019}. Additionally, we find that only a small fraction of the magnetar emission is transferred to the ejecta material, that is, $\xi{\sim}10^{-3}$.

The light curves of FXTs~15 and 20 show similar behavior and are well-fitted by plausible magnetar and ejecta parameters (see Table~\ref{tab:fitting_para}). Figures~\ref{fig:Ch4_posterior_trapped_FXT20_1}-\ref{fig:Ch4_posterior_trapped_FXT15_2} present their posterior distributions for nominal redshifts of $z{=}1.0$ and $0.5$, while 
Fig.~\ref{fig:fitting_trapped} shows their best fits and residuals. 
Unfortunately, due to the low number of bins for FXT~20 (dof${=}3$), parameters such as $\kappa$, $M_{\rm ej}$, $\xi$, and $P_i$ are not constrained properly. In particular, we find that the association of FXT~20 with a lanthanide-rich or free ejecta depends on the adopted redshift, that is, under the assumptions of $z{=}0.5$ and $1.0$ the opacities are $\kappa{<}40$ and ${\sim}0.1$~cm$^2$~g$^{-1}$, respectively. Similar to FXT~14, the magnetar luminosity contributes only a small fraction to the heating of the ejecta material (because of the low value of $\xi{\sim}10^{-3}$).
In the case of FXT~15, it has even fewer bins, requiring that during the fitting, we fixed the parameters $\kappa$ and $\xi$ (see Table~\ref{tab:fitting_para}). We consider only the lanthanide-free case based on the results of FXTs~14 and 20. Notably, FXT~15 has the most massive ejecta material (i.e., $M_{\rm ej}{\sim}10^{-3}$~M$_\odot$), and the highest $T_0$ (${\sim}10^2$~s). Both values are a consequence of the longer time to reach the peak luminosity, around ${\sim}10^3$~s, compared to FXTs~14 and 20 (${\lesssim}10^2$~s).

\begin{figure*}
    \centering
    \hspace{-0.5cm}
    \includegraphics[scale=1.0]{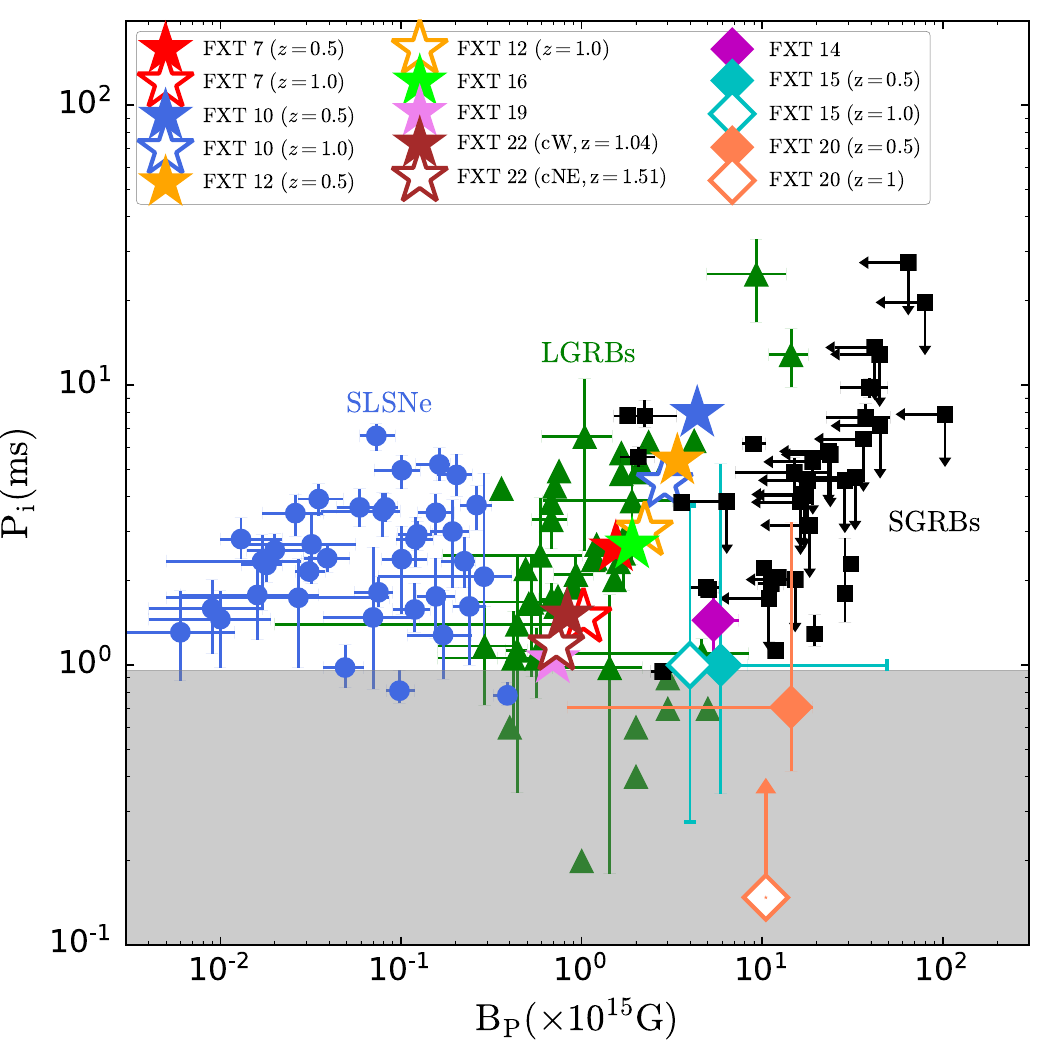}
    \vspace{-0.1cm}
    \caption{Comparison of the derived magnetar parameters in this work ($B_{\rm P}$ and $P_i$) with LGRBs \citep[green triangles;][]{Lyons2010,Yi2014}, SGRBs \citep[black squares;][]{Rowlinson2010,Rowlinson2013,Lu2015} and SLSNe \citep[blue circles;][]{Nicholl2017}.  The gray region represents the breakup limit of a neutron star, which is around 0.96 ms \citep{Lattimer2004}. FXTs related to the free and trapped models are represented by star and diamond markers, respectively.}
    \label{fig:magnetar_prop}
\end{figure*}

Within the magnetar scenario, the early rising phases in FXTs~14, 15, and 20 are interpreted as the emergence of the X-ray magnetar emission wind during its optically thick phase, while the declining phases of the light curves are well-fitted as the spin-down luminosity of the magnetar until the end of the detections. The best-fitted parameters suggest comparable magnetar parameters (magnetic field and initial rotational period values of ${\sim}10^{16}-10^{15}$~G and ${\sim}7.9$--$1.0$~ms, respectively). 

Finally, we must acknowledge the limitations of our modeling as far as these may influence interpretation. For instance, one weakness of our model is that it considers only a fixed opacity through the numerical solution. Yet, the only known example of a KN, AT2017gfo, exhibited two components with different opacities or multiple opacity compositions in the ejecta material. Clearly, AT2017gfo is a unique event, while a distribution of event properties from future KN detections is necessary to understand better the physics involved in these sources and the behavior of the opacity during and after the merger.

\section{Magnetar's parameters comparison}\label{sec:magnetar_result}

Around ${\approx}30-50$\% of SGRBs detected by \emph{Swift}-XRT have light curves that exhibit extended emission or an X-ray plateau compared to the standard afterglow power-law decay, which typically is interpreted as emission from a magnetar remnant \citep[see, for instance,][]{Rowlinson2013,Lu2015}. However, after the BNS merger, the young magnetar emission should be essentially isotropic, suggesting that a relatively large population of magnetar-driven FXTs without associated beamed gamma-ray emission should exist \citep[e.g.,][]{Zhang2013}. As such, we might expect that the magnetar parameters, $B_{\rm P}$ and $P_i$, for FXTs should be similar to those derived from the X-ray afterglows of SGRBs.

Figure~\ref{fig:magnetar_prop} compares the initial rotational period and magnetic field determined for our sample with long GRBs \citep[LGRBs;][]{Lyons2010,Yi2014}, SGRBs \citep{Rowlinson2013,Lu2015}, and super-luminous supernovae \citep[SLSNe;][]{Nicholl2017}. In general, SGRBs have the strongest magnetic fields ($B_P^{\rm SGRBs}{\lesssim}8{\times}10^{16}$~G) and a wide range of initial rotational periods ($P_i^{\rm SGRBs}{\lesssim}30$~ms); however, the majority of the magnetar parameters of SGRBs are just upper limits \citep[e.g.,][]{Lu2015}. On the other hand, SLSNe have the weakest magnetic fields ($B_P^{\rm SLSNe}{\lesssim}5{\times}10^{14}$~G) and a narrow range of initial rotational periods ($P_i^{\rm SLSNe}{\sim}8-1$~ms). LGRBs fall in parameter space between SGRBs and SLSNe, with $P_i^{\rm LGRBs}{\lesssim}20$~ms and $B_P^{\rm LGRBs}{\sim}5{\times}10^{15}-5{\times}10^{14}$~G, slightly overlapping with both kinds of objects.

Overall, the derived magnetar parameters from FXTs populate a region between LGRBs and SGRBs, overlapping the upper and lower portions of those populations, respectively. 
By contrast, the sample of FXTs and SLSNe do not overlap at all (refuting a possible association). This reinforces the results inferred from the host-galaxy properties presented in Papers I and II, and from the derived ejecta masses ($M_{\rm ej}{\sim}10^{-4}{-}10^{-2}$~M$_\odot$), which match well with simulations \citep{Hotokezaka2013,Bauswein2013,Sekiguchi2016,Ciolfi2017,Radice2018b,Siegel2019} and observations of GW~170817/AT2017gfo \citep{Abbott2017c}. FXTs related to the free and trapped zones appear to cover a different parameter-space region (see Fig.~\ref{fig:magnetar_prop}). The magnetar parameters derived for FXTs 15 and 20 carry large uncertainties, and as a result, they are consistent with both the free zone and trapped zone models (even considering the uncertainties related to redshifts and efficiency, which we discuss below). Only the well-constrained magnetar parameters of FXT~14 differ significantly from the free-zone FXTs (see Fig.~\ref{fig:magnetar_prop}).

In general, this sample of FXTs populates the lower tail of the magnetic field distribution of the magnetars suggested to drive SGRBs (i.e., ${\gtrsim}10^{15}$~G; see Fig.~\ref{fig:magnetar_prop}). Just a few outliers have ${\lesssim}10^{15}$~G (e.g., FXTs~19, and 22). Papers I and II suggest a non-association with on-axis LGRBs based on the lack of gamma-ray detections and the low luminosity of FXTs, although an association with low-luminosity LGRBs remains possible. 
The whole sample is above the breakup limit \citep[i.e., $P_i{\lesssim}1$~ms;][]{Lattimer2004} except for FXTs~15 and 20, although their parameters have large uncertainties or are only lower limits (such as FXT~20).

Because the magnetar model depends on the assumed efficiency, it is important to discuss the implications of the assumed value for $\eta$ during the comparison with other transients. As we explained above (see Sect.~\ref{sec:fitting}), we assumed an efficiency of $\eta{\sim}10^{-3}$ based on the previous analysis of FXTs CDF-S XT1 and XT2 \citep{Sun2019}; however, this parameter might take other values such as in previous works. For instance, \citet{Rowlinson2013} assumed an efficiency in the conversion of rotational energy into EM radiation of 100\%. In Appendix~\ref{sec:redshift_efficiency}, we discuss the changes in the magnetar parameters adopting a range of redshifts (especially relevant for the FXTs without measured redshifts) and efficiencies. Figure~\ref{fig:move} depicts the best-fit magnetar parameters under different redshifts and efficiencies. These higher efficiency values would increase our magnetar parameters by a factor between ${\approx}5$ and $7$, that is, rescaling our results in the right-top direction of Fig.~\ref{fig:magnetar_prop}. This displacement would mean that the FXT magnetar parameters overlap with those of SGRBs, implying that the magnetar nature of this sample of FXTs has similar parameters to those derived from the X-ray afterglows of SGRBs \citep[e.g.,][]{Rowlinson2010,Rowlinson2013,Gompertz2014,Lu2015}.

Finally, we estimate the volumetric rate (rescaling the values obtained by \citealt{Quirola2023} by a factor of 9/17) of the nine FXTs to ${\approx}$800--470~Gpc$^{-3}$~yr$^{-1}$. This value is consistent with those observed for GW~170817 \citep[$1.5_{-1.2}^{+3.2}{\times}10^3$~Gpc$^{-3}$~yr$^{-1}$;][]{Abbott2017b} and GRB~170817A \citep[$190_{-160}^{+440}$~Gpc$^{-3}$~yr$^{-1}$;][]{Zhang2018}. However, EM observations of GW~170817/GRB~170817A offer no conclusive evidence to rule out a long-lived or stable NS \citep[e.g.,][]{Abbott2017b,Piro2019,Troja2020}. The presence of a blue kilonova component related to a large mass of lanthanide-free ejecta and kinetic energy of $E_k{\approx}10^{51}$~erg, together with a successful relativistic jet, strongly disfavors a prompt collapse to a BH and suggests a first stage  HMNS formation that collapsed quickly to a BH after the merger \citep{Granot2017,Margalit2017,Shibata2017,Metzger2018,Rezzolla2018,Murguia_Berthier2021}. The uncertain nature of the remnant does not permit a direct comparison with FXTs.

Considering the possible BNS remnant channels of SGRBs, only the SMNS and stable-NS scenarios might be related to the FXT-magnetar sample.\footnote{The rate of BNS is $320^{+490}_{-240}$~Gpc$^{-3}$~yr$^{-1}$ \citep{Abbott2021a}} Under both scenarios, the percentage of mergers leading to the formation of SMNS and a stable NS are  ${\sim}$18--65\% and ${\lesssim}$3\%\footnote{Assuming the merging extragalactic NS-NS binary population is identical to the known Galactic NS--NS binaries.} \citep{Piro2017,Metzger2019,Margalit2019,Patricelli2020}, respectively, and the associated rates should be ${\sim}$530 and 30~Gpc$^{-3}$~yr$^{-1}$, respectively. Comparing both values and the FXT-magnetar sample, an association between the nine FXTs and the SMNS channel seems possible. Similarly, in the case of SGRBs \citep[the rate is ${\sim}$(2--0.1)${\times}10^3$~Gpc$^{-3}$~yr$^{-1}$ at $z{\lesssim}$1.5, after correcting by a nominal beaming factor of ${\approx}110$ and considering a Gaussian merger delay model;][]{Sun2015,Wanderman2015}, the rate of the SMNS channel is ${\sim}$1300--36~Gpc$^{-3}$~yr$^{-1}$, which is also consistent with this sub-sample of FXTs.

\begin{figure*}
    \centering
    \hspace{-0.5cm}
    \includegraphics[scale=0.72]{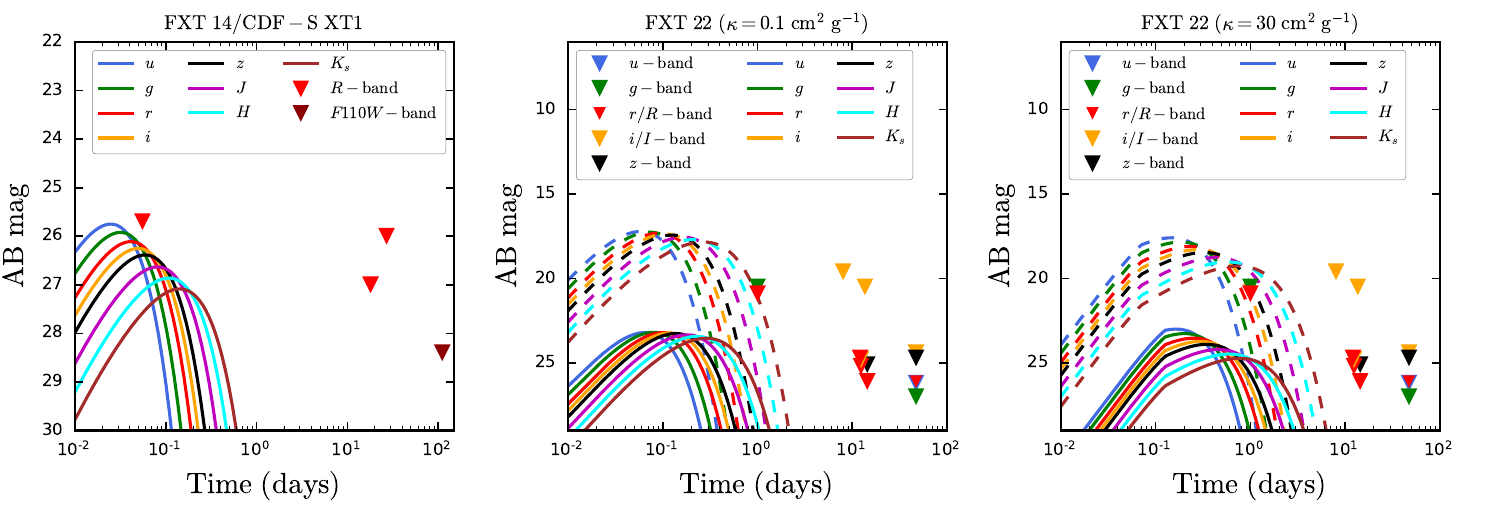}
    \caption{Numerical solution of the merger-nova emission in different energy bands. \emph{Left panel:} optical and NIR upper limits of FXT~14 and the expected merger-nova emission (from $u$- to $K$-bands) assuming their best fitting parameters (see Table~\ref{tab:fitting_para}) and $z=2.23$ (solid lines). As a  \emph{Middle and right panels:} optical and NIR upper limits of FXT~22, and the merger-nova emission considering the derived magnetar parameters (see Table~\ref{tab:fitting_para}), two redshifts of $z{=}1.0$ (solid lines) and $0.1$ (dashed lines), and assuming lanthanide-free (\emph{middle panel}; $\kappa{=}0.1$~cm$^2$~g$^{-1}$ and $M_{\rm ej}{=}10^{-4}$~M$_\odot$) and rich (\emph{right panel}; $\kappa{=}30$~cm$^2$~g$^{-1}$ and $M_{\rm ej}{=}10^{-3}$~M$_\odot$) scenarios. The upper limits of FXT~14 were taken from \citet{Bauer2017}, while for FXT~22 from \citet{Andreoni2021a,Andreoni2021b,Andreoni2021c}, \citet{Xin2021}, \citet{Rossi2021AT}, and \citet{Eappachen2023}. The time is measured from the X-ray trigger detection of the FXTs.}
    \label{fig:opt_NIR_UL}
\end{figure*}

Therefore, a possible association with off-axis SGRBs (due to the lack of a gamma-ray counterpart) and the BNS merger progenitor channel seems plausible, reinforcing previous conclusions drawn from the host-galaxy properties, lack of gamma-ray detections and low luminosities (Papers~I and II). The consistent range of FXT magnetar parameters also suggests a unified origin, with the distinguishing parameter being the viewing angle.

\section{Merger-nova emission in FXTs~14 and 22?}\label{sec:mergernova_FXTs}

Optical and NIR observations have been taken for FXTs~14 and 22 from tens of minutes to some days after the X-ray trigger, enabling constraints on contemporaneous counterparts. Unfortunately, however, no simultaneous counterparts have been identified \citep{Bauer2017,Jonker2021,Andreoni2021c}. Nevertheless, additional parameter space can be constrained from optical and NIR upper limits, provided the magnetar interpretation and merger-nova emission are plausible. Such contemporaneous observations are not available for the other FXTs, which we exclude from this line of discussion. In Sects.~\ref{sec:FXT_14} and \ref{sec:FXT_22}, we compare the most recent optical and NIR upper limits to the merger-nova theoretical models for FXTs~14 and 22, respectively.

\subsection{FXT~14}\label{sec:FXT_14}

FXT~14 was serendipitously discovered in the \emph{Chandra} Deep Field South (CDF-S) on 2014 October 01 \citep[2014-10-01 07:04:37 UT;][]{Luo2014,Bauer2017}. The CDF-S field has been extensively observed across the EM spectrum, and the position of FXT~14 was serendipitously observed ${\approx}80$ minutes after the X-ray trigger by the Visible Multi-Object Spectrograph (VIMOS) instrument mounted in the Very Large Telescope (VLT). No simultaneous optical transient was detected to an estimated magnitude limit of $m_R{\gtrsim}25.7$~mag at 2$\sigma$ confidence level. Subsequently, the field of the transient was imaged at ${\sim}$18, 27, and 111 days after the X-ray trigger by the VLT-FOcal Reducer/low dispersion Spectrograph 2 (FORS2), Gemini South-Gemini Multi-Object Spectrographs (GMOS), and the \emph{Hubble} Space Telescope (HST)-WFC3 instrument, respectively. Again, no optical or NIR counterpart was detected, yielding upper limits at the position of the transient of $m_R{\approx}27.0$, $m_r{\approx}26.0$, and $m_{\rm F110W}{\approx}28.4$~AB~mag at 2$\sigma$ confidence level for days ${\sim}$18, 27, and 111, respectively.

Using the best-fit parameters obtained for FXT~14 (see Table~\ref{tab:fitting_para}), we compute the numerical solution of the merger-nova emission in different energy bands (from Eq.~\ref{eq:024}). Figure~\ref{fig:opt_NIR_UL}, left panel, depicts the merger-nova emission model of FXT~14 considering different filters from $u$- to $K_s$-bands (solid color lines) and a comparison with the upper-limits of FXT~14 in the $R$, $r$ and F110W filters (F110W upper limit could be compared with $J$-band) adopting the redshift $z{=}2.23$. Overall, the upper limits remain consistent with the expected merger-nova theoretical model. Specifically, around the VLT-VIMOS observation (${\approx}80$~minutes or ${\approx}0.055$~ days after the trigger), we see that the upper limit still accommodates the peak emission in $r$-band from the numerical model for redshifts $z{\gtrsim}1.8$. Notably, this provides a more secure lower bound on the redshift of FXT~14 than the existing HST-based photometric redshift based on HST and deep ground-based imaging.

\subsection{FXT~22}\label{sec:FXT_22}

FXT~22 was serendipitously detected by a \emph{Chandra} calibration observation of Abell~1795 on 2021 April 23 \citep[2021-04-23 22:15:36.63 UT;][]{Lin2021}. No X-ray emission was detected in previous \emph{Chandra} observations at this location. Three extended optical objects have been identified as host galaxy candidates \citep[denoted cX, cNE, and cW;][]{Eappachen2023}. The transient location was imaged several times from ${\approx}1$ day to 45~days after the outburst by different telescopes and instruments.

The Zwicky Transient Facility (ZTF) observed the field of the transient three days before the outburst and one day after the X-ray trigger \citep{Andreoni2021a}. The forced photometry data, using the ZTF point-spread-function (PSF), was subtracted at the location of FXT~22. 
No source was detected in individual images in the $g$-, $r$-, and $i$-bands, just upper limits of $g{>}20.5$, $r{>}20.9$, and $i{>}19.6$~AB~mag ($5\sigma$ confidence value). 
Subsequently, the region was imaged by FORS2 on the 8m VLT telescope on 6 May 2021 (${\approx}13$ days after the X-ray trigger) in the $R$-band filter. No optical counterpart was found within the X-ray uncertainty region to a depth of $R{=}24.7$~mag at 3$\sigma$ confidence level \citep{Eappachen2023}.
The transient location was subsequently imaged with the Wafer-Scale Imager for Prime (WaSP) instrument mounted on the 200-inch Hale Telescope at Palomar Observatory \citep{Andreoni2021b} on 6 May 2021 at 07:10:00 UT in the $i$- and $r$-bands. No optical counterpart was identified, with upper limits of $i{>}24.8$ and $r{>}25.2$~mag.
On 7 May 2021, the field was observed by the NAOC 2.16m telescope at Xinglong Observatory with the $I$-band filter, yielding a non-detection with a limiting magnitude of $I{\gtrsim}20.5$~mag in stacked image \citep{Xin2021}. 
On 8 May 2021 at 07:28:00 UT (i.e., ${\approx}15$ days after the transient detection), the field was visited by the 8.4m Large Binocular Telescope (LBT) using the $r$-sloan and $z$-sloan bands with the Large Binocular Cameras (LBC) \citep{Rossi2021AT}. No detection at the position of the transients was identified, with limit magnitudes of $r{>}26.1$ and $z{>}25.1$~mag. 
Finally, on 10 June 2021 (i.e., ${\approx}45$ days after the onset of the outburst), the quintuple-beam imager HiPERCAM instrument mounted on the 10.4~m Gran Telescopio Canarias (GTC) observed the field of the source with simultaneous $u_s$, $g_s$, $r_s$, $i_s$, and $z_s$-band filters. No detection at the position of the X-ray transients was identified, with limiting magnitudes of $u_s{>}26.2$, $g_s{>}27.0$, $r_s{>}26.1$, $i_s{>}24.4$, and $z_s{>}24.7$~mag at 3$\sigma$ confidence level \citep{Eappachen2023}.

As noted in Sect.~\ref{sec:trapped_zone}, the difference between the free- and trapped zones is geometrical (see Fig.\ref{fig:cartoon}), and events such as FXT~22 (which is related to the free zone emission based on its light curve plateau) could ultimately be associated with a line of sight that passed through the trapped zone where the ejecta mass was relatively low. Under this condition, FXT~22 would follow the relation $t_\tau{\lesssim}T_X$ (i.e., the time at which point the ejecta material becomes optically thin occurs before the plateau appears). In this way, we explore in Fig.~\ref{fig:opt_NIR_UL} whether the merger-nova models, assuming either the lanthanide-free ($\kappa{=}0.1$~cm$^2$~g$^{-1}$ and $M_{\rm ej}{=}10^{-4}$~M$_\odot$) or -rich ($\kappa{=}30$~cm$^2$~g$^{-1}$ and $M_{\rm ej}{=}10^{-3}$~M$_\odot$) scenarios, are consistent with the optical upper limits obtained for FXT~22.
We find that the upper limits remain consistent with the expected merger-nova theoretical model for redshifts $z{\gtrsim}0.1$ (see Fig.~\ref{fig:opt_NIR_UL}, middle and right panels, dashed lines), and hence remain completely consistent with potential hosts cW at $z_{cW}{=}1,04$ and cNE at $z_{cNE}{\approx}1.5$. This limit also implies that the strongest host candidate, the faint source cX, visible in GTC-HiPERCAM $g$-filter images only  \citep{Eappachen2023}, should lie at $z_{\rm cX}{\gtrsim}0.1$.

\begin{table}
    \centering
    \caption{The parameters of various NS EoS models \citep{Lasky2014,Ravi2014,Li2016,Ai2018}.}
    \hspace{-0.5cm}
    \scalebox{0.82}{
    \begin{tabular}{cccccc}
    \hline \hline
    EoS & $M_{\rm TOV}$($M_\odot$) & $R_{\rm M}$(km) & $I$(10$^{45}$~g~cm$^2$) &  $\hat{\alpha}$(10$^{-10}$~s$^{-\hat{\beta}}$) & $\hat{\beta}$  \\ 
    (1) & (2) & (3) & (4) & (5) & (6) \\ \hline
    SLy & 2.05 & 9.99 & 1.91 & 1.60 & $-2.75$ \\
    BSk20 & 2.17 & 10.17 & 3.50 & 3.39 & $-2.68$ \\
    APR & 2.2 & 10 & 2.13 & 0.303 & $-2.95$ \\
    GM1 & 2.37 & 12.05 & 3.33 & 1.58 & $-2.84$ \\
    DDME2 & 2.48 & 12.09 & 5.85 & 1.966 & $-2.84$ \\
    AB-N & 2.67 & 12.90 & 4.3 & 0.112 & $-3.22$ \\
    AB-L & 2.71 & 13.70 & 4.7 & 2.92 & $-2.82$ \\
    NL3$\omega\rho$ & 2.75 & 12.99 & 7.89 & 1.706 & $-2.88$ \\
\hline
    \end{tabular}}
    \tablefoot{\emph{Column 1:} EoS name. \emph{Column 2:} maximum mass for a non-rotating NS or Tolman-Oppenheimer-Volkoff (TOV) mass. \emph{Columns 3 and 4:} Radius and inertia of the NSs, respectively. \emph{Columns 5 and 6:} parameters of the NS for each EoS, which are related to the maximum gravitational mass of a rotating NS as $M_{\rm max}{=}M_{\rm TOV}(1{+}\hat{\alpha} P^{\hat{\beta}})$, where $P$ is the rotational period.}
    \label{tab:EoS_para}
\end{table}

\begin{figure*}
    \centering
    \includegraphics[scale=0.8]{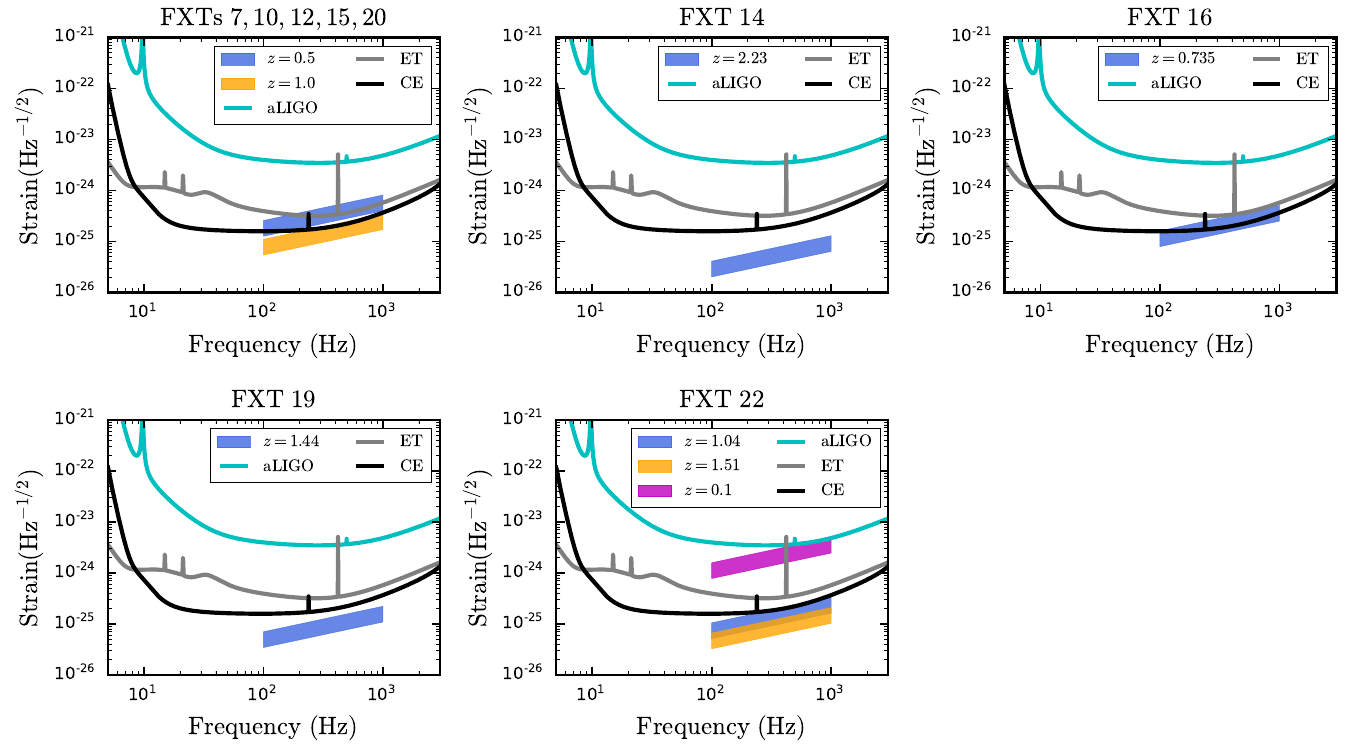}
    \caption{GW strain evolution with frequency for the FXTs analyzed in this work. Each shaded region represents the GW strain considering eight EoS (see Table~\ref{tab:EoS_para}) at different distances. The cyan, gray, and black lines are the sensitivity limits for aLIGO, ET, and CE GW telescopes, respectively. For the FXTs without redshifts (FXTs~7, 10,12, 15, and 20), we assumed $z{=}0.5$ and $1.0$ (see Table~\ref{tab:fitting_para}).}
    \label{fig:strain_gw}
\end{figure*}

\section{Rotational energy losses via GWs and detection probability}\label{sec:GW_emission}

A fraction of the rotational energy of the magnetar is likely released by GW radiation \citep{Fan2013,Lasky2014,Lan2020}, produced via mass quadrupole deformation (related to the ellipticity, $\epsilon$) considering rigid-body rotation \citep{Zhang2001,Fan2013,Ho2016,Lasky2016,Lu2017}. There are several possibilities for deforming a newly born magnetar \citep[e.g., a bar-mode instability or inertial quadrupole $r$-mode;][]{Andersson2001,Cutler2002,Haskell2008,Corsi2009,Lasky2016}. Nevertheless, the most important contribution to $\epsilon$ may come from magnetic induction. For a differentially rotating magnetar, its magnetic fields become twisted and stored in a toroidal shape, causing it becomes non-spherical \citep{Cutler2002,Haskell2008}. The condition to produce significant GW radiation is for the magnetic axis to be strongly misaligned with the rotational axis \citep{Shapiro1983,Zhang_book_2018}.

We briefly explore a potential GW emission contribution in the context of the magnetar model. As we explained above, the millisecond magnetar also loses rotational energy via GW emission,
where the characteristic spin-down timescale (Eq.~\ref{eq:005}) of the magnetar can be numerically written as
\begin{equation}
    \tau_{\rm GW}=\frac{5c^5P_i^4}{128GI\epsilon^2(2\pi)^4}{\simeq}9.1{\times}10^{3}\left(\frac{I}{10^{45}}\right)^{-1}\left(\frac{\epsilon}{10^{-3}}\right)^{-2}\left(\frac{P_i}{10^{-3}}\right)^4~\text{s}.
    \label{eq:026}
\end{equation}
Given that the decay phase of the FXTs remains inconsistent with a GW dominant spin-down luminosity trend (see Eq.~\ref{eq:007}), we expect that $\tau_{\rm GW}{>}T_X$ (where $T_X{\gtrsim}1$~ks), and combined with Eq.~\ref{eq:026}, an upper limit on the ellipticity can be expressed as
\begin{equation}
    \epsilon{<}2.5{\times}10^{-3}\left(\frac{I}{10^{45}}\right)^{-1/2}\left(\frac{P_i}{10^{-3}}\right)^2.
    \label{eq:027}
\end{equation}
To compute the maximum value of $\epsilon$, we should take into account different EoSs (see Table~\ref{tab:EoS_para}). Based on these, the maximum values of $\epsilon$, assuming $P_i{\approx}1$~ms (see Table~\ref{tab:fitting_para}), lie in the range  $\epsilon{\lesssim}1.2{\times}10^{-3}-2.1{\times}10^{-4}$. This range is lower than the expected values when GW radiation dominates the rotational energy losses \citep[i.e., $\epsilon{\sim}10^{-2}$;][]{Fan2013,Lasky2014,Ho2016,Lan2020}, 
reinforcing the assumed value of ${\epsilon}{\sim}10^{-5}$ during the fitting process.

If this new population of magnetars does lose energy via GWs, it should be possible to constrain their GW emission considering a set of EoSs (see Table~\ref{tab:EoS_para}), as well as predict their detectability with the current Advanced LIGO detector (aLIGO) and more sensitive future detectors such as the \emph{Einstein Telescope} (ET) and \emph{Cosmic Explorer} (CE). Specifically, if the rotation energy is released via GWs with a frequency $f$, the GW strain ($h(t)$) for a rotating NS at luminosity distance $d_L$ can be expressed as \citep[][and references therein]{Lu2017,Lu2019}
\begin{equation}
    h(t)=\frac{4GI\epsilon}{d_L c^4}\Omega(t)^2,
    \label{eq:028}
\end{equation}
where the characteristic amplitude of GWs from a rotating NS can be estimated as \citep{Corsi2009,Hild2011,Lasky2016,Lu2017}
\begin{equation}
\begin{split}
h_c&=fh(t)\sqrt{\frac{dt}{df}}=\frac{f}{d_L}\sqrt{\frac{5GI}{2c^3f}}\\
&{\simeq}8.22\times10^{-24}\left(\frac{I}{10^{45}~{\rm g~cm^{2}}}\right)^{1/2}\left(\frac{f}{1000~{\rm Hz}}\right)^{1/2}\left(\frac{d_{L}}{100~{\rm Mpc}}\right)^{-1},
\end{split}
    \label{eq:029}
\end{equation}
where $f$ is the frequency in Hz and $d_{L}$ is in Mpc, and the gravitational wave frequency evolution, $df/dt$, is derived directly from Eq.~\ref{eq:002}. It is clear that $h_c$ does not depend on $\epsilon$, just $d_L$ and $I$. Thus, each EoS will provide a different characteristic amplitude. It is important to emphasize that future GW detectors should also detect the GWs from the BNS merger, which precedes the FXT magnetars.

Figure~\ref{fig:strain_gw} compares the expected $h_c$ considering eight EoSs from Table~\ref{tab:EoS_para} (shaded regions) for the FXTs analyzed in this work with the sensitivities of aLIGO (cyan line), ET (gray line), and CE (black line). 
Overall, the maximum strain values, $h_c$, are for the EoS NL3$\omega\rho$, which has the highest inertia (see Table~\ref{tab:EoS_para}). All of the FXTs remain firmly beyond the current capabilities of aLIGO (i.e., $z{\lesssim}$0.03 or ${\lesssim}$130~Mpc). ET will be sensitive to events such as FXTs~7, 10, 12, 15, and 20 if they lie at redshifts $z{\lesssim}0.5$, while CE should push out to $z{\lesssim}0.75$ (allowing the inclusion of events like FXT~16). No GW observatory, existing or currently planned, will be able to detect events similar to FXTs~14, 19, and 22 (see Fig.~\ref{fig:strain_gw}).

\section{Conclusions} \label{sec:conclusion}

Papers I and II reported the detection of 22 extragalactic fast X-ray transients (FXTs) hidden in two decades of \emph{Chandra} data (analyzing ${\sim}$259~Ms of data), among which 17 are associated with distant galaxies (${\gtrsim}$100~Mpc). Different mechanisms and progenitors have been proposed to explain their properties; nevertheless, their nature remains uncertain.
We interpreted a subset of nine distant FXTs within the framework of a  two-zone ('free' and 'trapped') BNS merger magnetar model, parametrized by a magnetic field ($B_{\rm P}$), initial rotational period ($P_i$), ejecta mass ($M_{\rm ej}$), opacity ($\kappa$), and the fraction of the magnetar energy that is transferred to the heating of the ejecta ($\xi$). 

The model can explain well the observed X-ray light curves of the sources, wherein we associate six FXTs to a free-zone scenario and three FXTs to a trapped-zone scenario. The best-fitted results yield relatively similar $B_{\rm P}$ and $P_i$ values in the ranges $10^{16}$--$10^{15}$~G and $7.9$--$1.0$~ms, respectively. These values are comparable to those derived from the magnetar interpretation of some X-ray afterglow of SGRBs \citep[for instance][]{Rowlinson2010,Rowlinson2013,Lu2015}, reinforcing an association with off-axis SGRBs as suggested previously \citep{Bauer2017,Xue2019,Lin2022,Quirola2022,Quirola2023}.

For two FXTs, XRT~141001 (FXT~14) and XRT~210423 (FXT~22), which have prompt deep optical and near-infrared upper limits available, we explored the constraints on early merger-nova detections. We found that the merger-nova numerical models agree with the optical and near-infrared upper limits, and we identified that the non-detections place lower limits on the distance of $z{\gtrsim}1.8$ and ${\gtrsim}0.1$ for XRT~141001/FXT~14 and XRT~210423/FXT~22, respectively.

Finally, we analyzed the contribution of the gravitational wave (GW) radiation to the magnetar spin-down (related to the ellipticity factor, $\epsilon$, see Eq.~\ref{eq:002}), and the detectability of GWs associated with our sample of FXTs by current and future GW observatories. The maximum value of $\epsilon$, considering different equations of state models (see Table~\ref{tab:EoS_para}), is $\epsilon{\lesssim}10^{-3}-10^{-4}$, which is below the expected values for GW losses to dominate the system \citep[i.e., $\epsilon{\gtrsim}10^{-2}$;][]{Fan2013,Lasky2014,Ho2016,Lan2020}. We find that it is only possible to detect BNS magnetar remnants similar to the FXTs with the current Advanced LIGO detectors to $z{\lesssim}0.03$, while future GW detectors should extend our sensitivity to $z{\lesssim}0.5$--0.75.

\section*{Acknowledgements}

The scientific results reported in this article are based on observations made by the {\it Chandra} X-ray Observatory. This research has made use of software provided by the {\it Chandra} X-ray Center (CXC). We acknowledge support from: CONICYT through Programa de Capital Humano Avanzado, folio \#21180886 (JQ-V) Basal AFB-170002 (JQ-V, FEB), and FONDECYT Regular 1141218 (JQ-V, FEB); the Ministry of Economy, Development, and Tourism's Millennium Science Initiative through grant IC120009, awarded to The Millennium Institute of Astrophysics, MAS (JQ-V, FEB); this project was (partially) funded by NWO under grant number 184.034.002 (P.G.J.); GY acknowledges funding from the Netherlands Research School for Astronomy (NOVA).
NSF grant AST-2106990 and \emph{Chandra} X-ray Center grant GO0-21080X (W.N.B.); the National Natural Science Foundation of China grant 11991053 (B.L.); the NSFC grant (12025303), the Strategic Priority Research Program of the Chinese Academy of Sciences (grant NO. XDB0550300; Y.Q.X.); H.S. is supported by the National Natural Science Foundation of China grant 12103065.




\bibliographystyle{mnras}
\bibliography{Quirola-Vasquez2019}



\appendix

\section{the magnetar model in detail} \label{sec:theoretical_detail}

Here, we consider the formalism from \citet{Yu2013} and \citet{Sun2017} to describe the physical model of the magnetar emission. A nutshell description was done in Sect.~\ref{sec:theoretical}.

\subsection{Free-zone emission}

The total rotational energy reservoir of a millisecond magnetar is
\begin{equation}
  E_{\rm rot}=\frac{1}{2}I\Omega^2,
  \label{eq:001}
\end{equation}
where $I$ is the moment of inertia, $\Omega{=}2\pi/P$ is the angular frequency of the magnetar, and $P$ its period. The millisecond magnetar loses its rotational energy through electromagnetic (EM) and gravitational-wave (GW) radiation \citep[from non-axisymmetric deformation;][]{Shapiro1983,Usov1992,Zhang2001,Gao2016,Lasky2016,Sun2017} as:
\begin{equation}
    \dot{E}_{\rm rot}=I\Omega\dot{\Omega}=\dot{E}_{\rm EM}+\dot{E}_{\rm GW}=-\frac{B_p^2R_{\rm M}^6\Omega^4}{6c^3}-\frac{32GI^2\epsilon^2\Omega^6}{5c^5},
    \label{eq:002}
\end{equation}
where $\dot{\Omega}$ is the time derivative of the angular frequency, $B_{\rm P}$ is the dipole component of the magnetic field at the poles, $\epsilon$ is the ellipticity of the magnetar, $R_{\rm M}$ is the magnetar radius, and $c$ and $G$ are the speed of light and Newton gravitational constant, respectively. 

The evolution of $\Omega(t)$ considers the EM and GW losses, which both play a role in the spin-down evolution. In the following, we define the dipole spin-down luminosity as \citep{Zhang2001,Metzger2014,Lasky2016,Sun2017,Zhang_book_2018}:
\begin{equation}
    L_{\rm sd}(t)=\frac{B_P^2R_{\rm M}^6\Omega(t)^4}{6c^3},
\end{equation}
where $\Omega(t)$ is the solution of the previous spin-down Equation~\ref{eq:002}. 
Importantly, a strong magnetic field provides a mechanism for extracting rotational energy from the NS remnant via EM spin-down. Indeed, magneto-hydrodynamic simulations show that the magnetic fields in a BNS merger are amplified to values exceeding the field of Galactic magnetars \citep[i.e., $B_P{\sim}10^{15}-10^{16}$~G;][]{Price2006,Zrake2013,Kiuchi2014}.
Discarding one of the two terms in Eq.~\ref{eq:002}, assuming the other is dominant, we can define the EM and GW characteristic spin-down timescales as $\tau_{\rm EM}$ and $\tau_{\rm GW}$, respectively, \citep{Zhang2001}:
\begin{equation}
    \tau_{\rm EM}=\frac{3c^3I}{B_P^2R_{\rm M}^6\Omega_i^2}=\frac{3c^3IP_i^2}{B_P^2R_{\rm M}^6(2\pi)^2}~{\rm s},
    \label{eq:004}
\end{equation}
\begin{equation}
    \tau_{\rm GW}=\frac{5c^5}{128GI\epsilon^2\Omega_i^4}=\frac{5c^5P_i^4}{128GI\epsilon^2(2\pi)^4}~{\rm s},
    \label{eq:005}
\end{equation}
where $\Omega_i$ and $P_i$ are the initial angular velocity and period of the magnetar, respectively. On times shorter than the spin-down timescale (called $t_{\rm sd}$, and  defined as $t_{\rm sd}{=}\min[\tau_{\rm EM},\tau_{\rm GW}]$), it is expected that $\Omega(t)$ and $L_{\rm sd}(t)$ will remain stable. 
Considering the EM spin-down term only in Eq.~\ref{eq:002}, the luminosity evolves as
\begin{equation}
    L_{\rm sd}(t)=L_{0}^{\rm EM}\left(1+\frac{t}{\tau_{\rm EM}}\right)^{-2}~{\rm erg/s},
    \label{eq:006}
\end{equation}
where $L_{0}^{\rm EM}{=}I\Omega_i^2(2\tau_{\rm EM})^{-1}$~erg~s$^{-1}$. Eq.~\ref{eq:006} implies that the light curve shows a plateau whereby $L_{\rm sd}{\propto}L_{0}^{\rm EM}$ at $t{\lesssim}\tau_{\rm EM}$, followed by a decay $L_{\rm sd}{\propto}L_{0}^{\rm EM}(t/\tau_{\rm EM})^{-2}$ at $t{\gg}\tau_{\rm EM}$.

The above equations, however, neglect the enhanced angular momentum losses due to neutrino-driven mass loss, which are important at early times during the magnetar formation \citep{Metzger2011}.
On the other hand, considering only the GW spin-down term in Eq.~\ref{eq:002} (i.e., the angular velocity evolution is mainly due to the GW losses), the spin-down luminosity behaves as
\begin{equation}
    L_{\rm sd}(t)=L_{0}^{\rm EM}\left(1+\frac{t}{\tau_{\rm GW}}\right)^{-1}~{\rm erg/s}.
    \label{eq:007}
\end{equation}
Similar to Eq.~\ref{eq:006}, the light curve will show a plateau $L_{\rm sd}{\propto}L_{0}^{\rm EM}$ at $t{\lesssim}\tau_{\rm GW}$, followed by a decay as $L_{\rm sd}{\propto}L_{0}^{\rm EM}(t/\tau_{\rm GW})^{-1}$ at $t{\gg}\tau_{\rm GW}$. 
Such rapid rotation can produce large deformations, which imply significant quadrupole GW radiation, assuming rigid body rotation. It is important to note that the luminosity decays at late times as $L_{\rm sd}(t){\propto}t^{-1}$ (Eq.~\ref{eq:007}), rather than $L_{\rm sd}(t){\propto}t^{-2}$ (Eq.~\ref{eq:006}) like in the EM dipole spin-down scenario.

In reality, if GW radiation is to dominate spin-down, it must do so early in the evolution, while EM emission becomes important later \citep{Lasky2016}.
The conditions for this come from the transition timescale once both the EM and GW terms are considered, and the spin-down luminosity behaves as
\begin{equation}
    L_{\rm sd}(t){\propto}\left\{ 
    \begin{matrix}
    t^0 & t{<}t_{\rm sd}=\tau_{\rm GW} \\
    t^{-1} & t_{\rm sd}=\tau_{\rm GW}{<}t{<}\tau_{\rm EM} \\
    t^{-2} & t{>}\tau_{\rm EM}
    \end{matrix}
    \right.,
    \label{eq:008}
\end{equation}
for $\tau_{\rm GW}{<}\tau_{\rm EM}$, and
\begin{equation}
    L_{\rm sd}(t){\propto}\left\{ 
    \begin{matrix}
    t^0 & t{<}t_{\rm sd}=\tau_{\rm EM} \\
    t^{-2} & t{>}t_{\rm sd}
    \end{matrix}
    \right.,
    \label{eq:009}
\end{equation}
for $\tau_{\rm EM}{<}\tau_{\rm GW}$.
It remains unclear which is the preferential regime to lose rotational kinetic energy. To acquire evidence, it is necessary to detect a substantial number of new GW events and their associated EM signals using GW detectors and high-energy instruments, respectively, in the future \citep{Lu2018}.


\subsection{Trapped-zone emission}

The total energy of the ejecta (excluding the rest mass energy) can be expressed as
\begin{equation}
    E_{\rm ej}=(\Gamma-1)M_{\rm ej}c^2+\Gamma E_{\rm int}^\prime,
    \label{eq:012}
\end{equation}
where $E_{\rm int}^\prime$ is the internal energy in the co-moving frame\footnote{The parameters in the co-moving frames are denoted as $Q^\prime$. Parameters without a prime ($^\prime$) are in the observer frame.}. The first and second terms represent the kinetic minus the rest-mass energy and the internal energy of the ejecta, respectively. At each time step, $dt$, the ejecta material is boosted by the energy of the magnetar ($L_{\rm sd}$) and the radioactive decay of the $r$-process material ($L_{\rm ra}$). Moreover, the bolometric luminosity of the heated electrons is a lost energy source ($L_e$) in the system. From energy conservation, we can conclude that in the observer frame
\begin{equation}
    dE_{\rm ej}=(L_{\rm sd}+L_{\rm ra}-L_e)dt.
    \label{eq:013}
\end{equation}

The conversion from observer time ($t$) to co-moving time ($t^\prime$) is related as $dt^\prime=\mathcal{D}dt$, where $\mathcal{D}{=}1/[\Gamma(1-\beta\cos\theta)]$ is the Doppler factor with $\beta{=}\sqrt{1-\Gamma^{-2}}$ and $\theta{=}0^\circ$ for an on-beam observer. Deriving Eq.~\ref{eq:012} in terms of $dt$ and equating to Eq.~\ref{eq:013}, we obtain the evolution of the Lorentz factor as a function of the observer time as
\begin{equation}
    \frac{d\Gamma}{dt}=\frac{L_{\rm sd}+L_{\rm ra}-L_e-\Gamma\mathcal{D}(dE_{\rm int}^\prime/dt^\prime)}{M_{\rm ej}c^2+E_{\rm int}^\prime}.
\end{equation}
In the last equation, we can see that the evolution of $\Gamma$ also depends on the change of the internal energy of the ejecta material in the co-moving frame ($dE_{\rm int}^\prime/dt^\prime$). To convert the luminosities from the observer to the co-moving frames, we use the relation $L^\prime=L/\mathcal{D}^2$ \citep{Zhang_book_2018}.

The radioactive luminosity in the co-moving frame is described as \citep{Korobkin2012}
\begin{equation}
    L_{\rm ra}^\prime=4\times10^{51}\left(\frac{M_{\rm ej}}{M_\odot}\right)\times\left[\frac{1}{2}-\frac{1}{\pi}\arctan\left(\frac{t^\prime-t_0^\prime}{t_\sigma^\prime}\right)\right]^{1.3}~{\rm erg}~{\rm s}^{-1},
    \label{eq:016}
\end{equation}
where $t_0^\prime{\sim}$1.3~seconds and $t_\sigma^\prime{\sim}$0.11 seconds are the best-fitted parameters of the nuclear heating power as a function of time from simulations of heavy element
nucleosynthesis in compact binary mergers \citep{Korobkin2012}.
In the case of a relativistic gas, the pressure (dominated by radiation) in the co-moving frame can be written as a function of the co-moving internal energy density as
\begin{equation}
    P^\prime=\frac{1}{3}\frac{E_{\rm int}^\prime}{V^\prime}.
    \label{eq:017}
\end{equation}
The co-moving volume evolution is determined by
\begin{equation}
    \frac{dV^\prime}{dt^\prime}=4\pi R_{\rm ej}^2\beta c,
    \label{eq:018}
\end{equation}
while the radius evolution of the ejecta material in the observer frame can be described as
\begin{equation}
    \frac{dR_{\rm ej}}{dt}=\frac{\beta c}{1-\beta}.
    \label{eq:019}
\end{equation}

Additionally, as we explain above, the ejecta suffers a cooling via a bolometric electron emission. The co-moving frame radiated bolometric luminosity, which depends on the ejecta optical depth and the diffusion approximation, can be estimated as \citep{Kasen2010,Kotera2013}
\begin{equation}
    L_e^\prime=\left\{
    \begin{matrix}
    \frac{E_{\rm int}^\prime c\Gamma}{\tau R_{\rm ej}} & {\rm for}\ t{<}t_\tau \\
     & \\
    \frac{E_{\rm int}^\prime c\Gamma}{R_{\rm ej}} & {\rm for}\ t{\gtrsim}t_\tau
    \end{matrix}
    \right. ,
    \label{eq:020}
\end{equation}
where the term for $t{<}t_\tau$ takes into account the skin-depth effect of an optically thick emitter \citep{Yu2013}, and $t_\tau$ is the time when $\tau{=}1$.

The spectrum of the merger nova should approximately resemble a blackbody under the co-moving temperature, $T^\prime$, and the peak energy of the blackbody emission is:
\begin{equation}
    \varepsilon_{\gamma,p}\approx4\mathcal{D}k_bT^\prime=\left\{
    \begin{matrix}
    4\mathcal{D}k_b\left(\frac{E_{\rm int}^\prime}{aV^\prime\tau}\right)^{1/4} & {\rm for}\ \tau{>}1 \\
     & \\
    4\mathcal{D}k_b\left(\frac{E_{\rm int}^\prime}{aV^\prime}\right)^{1/4} & {\rm for}\ \tau{\lesssim}1 \\
    \end{matrix}
    \right. ,
\end{equation}
where $k_b$ and $a$ are the Boltzmann and the blackbody radiation constants, respectively.

\section{role of redshift and efficiency factor} \label{sec:redshift_efficiency}

\begin{figure}
    \centering
    \includegraphics[scale=0.75]{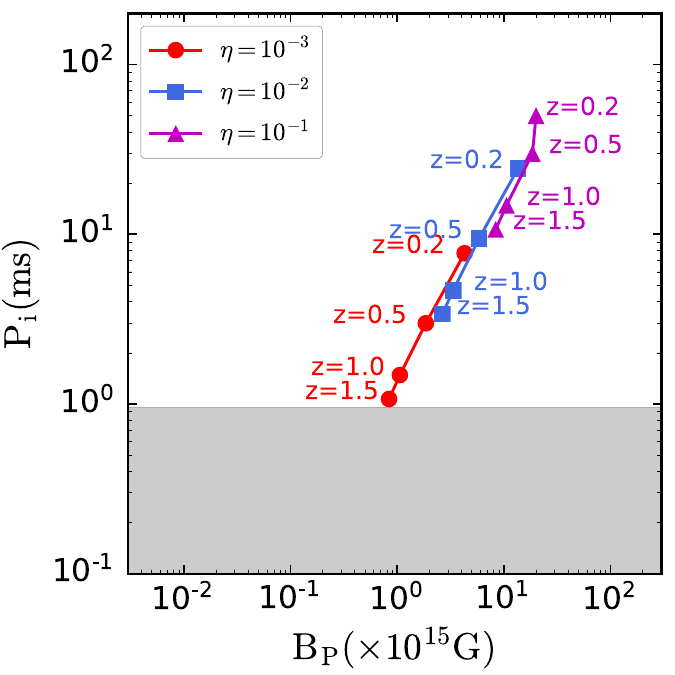}
    \vspace{-0.3cm}
    \caption{This graph depicts the best-fit magnetar parameters at different redshifts and efficiencies. The number beside each data point is the corresponding redshift, while the colors represent efficiencies of $\eta{\sim}10^{-3}$ (red lines), $10^{-2}$ (blue lines), and $10^{-1}$ (magenta lines). The gray region represents the breakup limit of a neutron star, which is $\approx$0.96 ms \citep{Lattimer2004}.}
    \label{fig:move}
\end{figure}

As mentioned above, the redshift uncertainties affect the derived magnetar parameters. We recomputed the magnetar parameters adopting a range of redshifts (from $z{=}0.2$ to $1.5$) to explore this change. In Figure~\ref{fig:move}, we can see how the magnetic field and the initial period move over a wide redshift range. Under a redshift modification from $z{=}$0.2 to 1.0, both the initial period and magnetic field decrease by a factor of ${\approx}7$ and $5$, respectively (see Fig.~\ref{fig:move}). 

Moreover, during the fitting process (see Sect.~\ref{sec:fitting}), we fix the efficiency factor to $\eta{\sim}10^{-3}$ based on literature outcomes \citep{Xue2019}. Here, we analyze its role in the magnetar parameters modifying the efficiency factor in Fig.~\ref{fig:move} (from $\eta{=}10^{-3}$ to $10^{-1}$). It is clear that the magnetic field and the initial rotational period increase by a factor of ${\approx}5$ and $7$, respectively, where $\eta$ is increased from $10^{-3}$ to $10^{-1}$. A similar factor has been found in previous works \citep[e.g.,][]{Rowlinson2010,Rowlinson2013}.

\section{posterior distributions for the free and trapped zones}

\begin{figure*}
    \hspace{-1.5cm}
    \centering
    \includegraphics[scale=0.55]{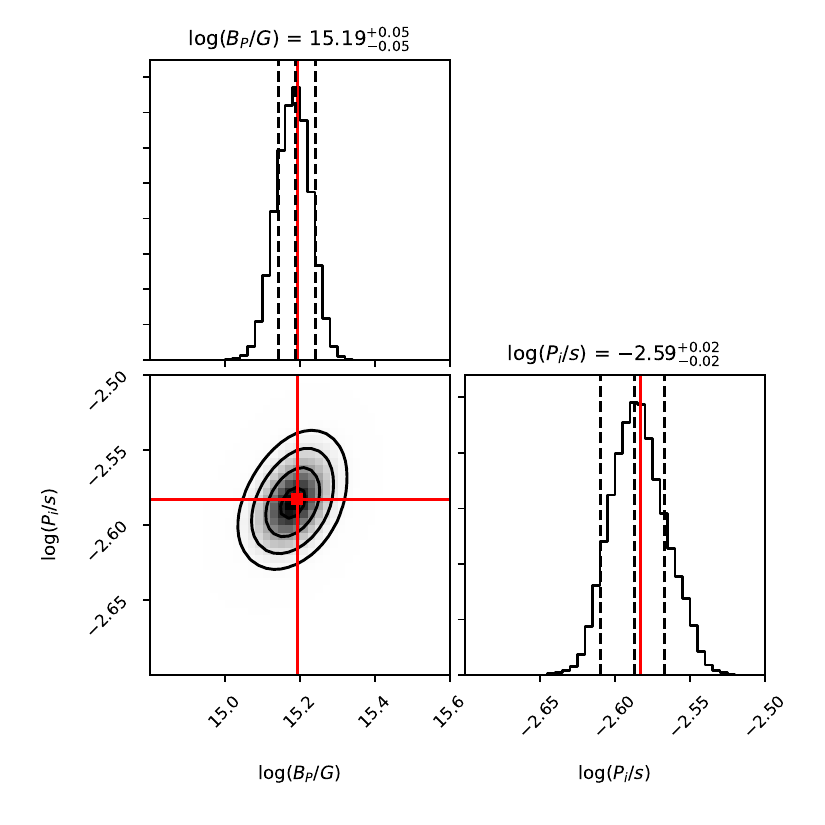}
    \includegraphics[scale=0.55]{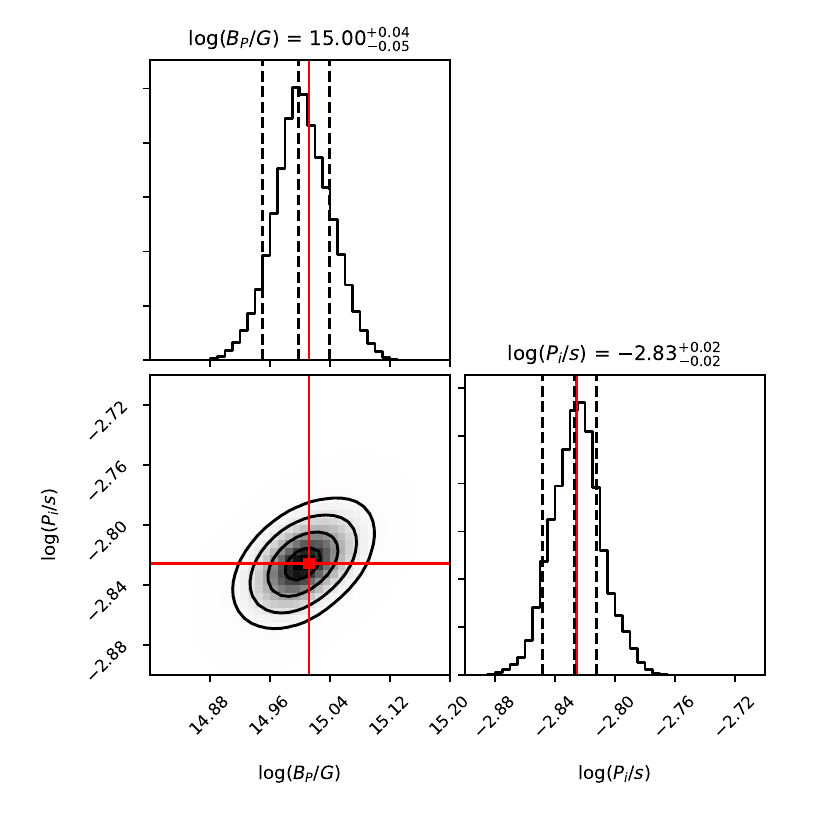}
    \vspace{-0.3cm}
    \caption{MCMC posterior distribution for the free zone fitting result of FXT~7 assuming $z_{\rm nominal}{=}0.5$ (left panel) and $1.0$ (right panel). The dashed black lines show the 16\%, 50\% and 84\% percentile of the MCMC results, while the red lines depicts the parameters of the best-fitted model.}
    \label{fig:Ch4_posterior_free_FXT7}
\end{figure*}

\begin{figure*}
    \hspace{-1.5cm}
    \centering
    \includegraphics[scale=0.55]{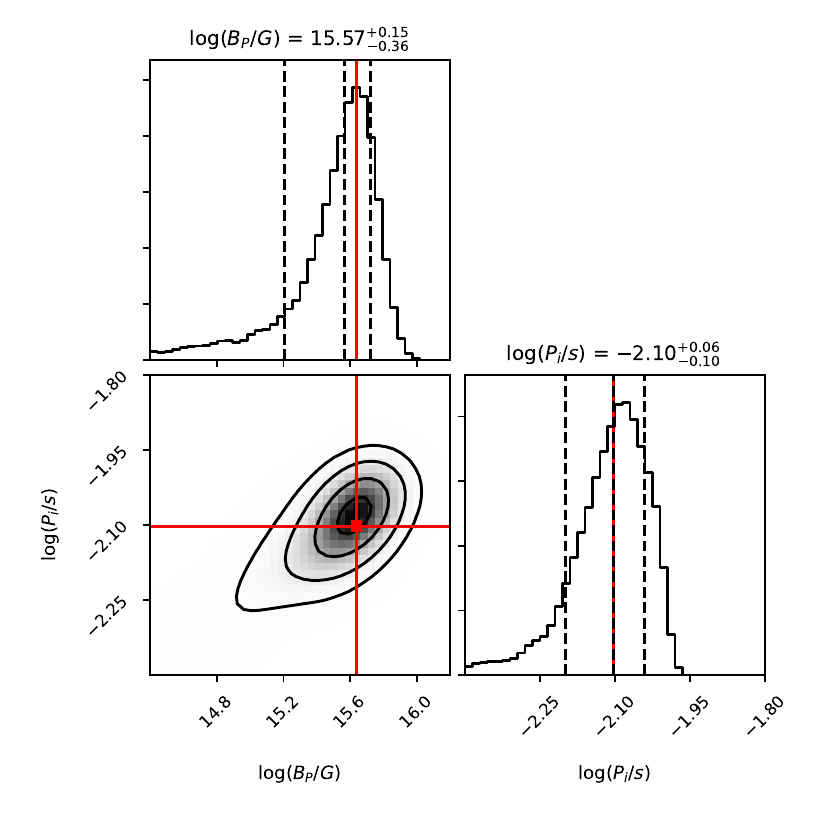}
    \includegraphics[scale=0.55]{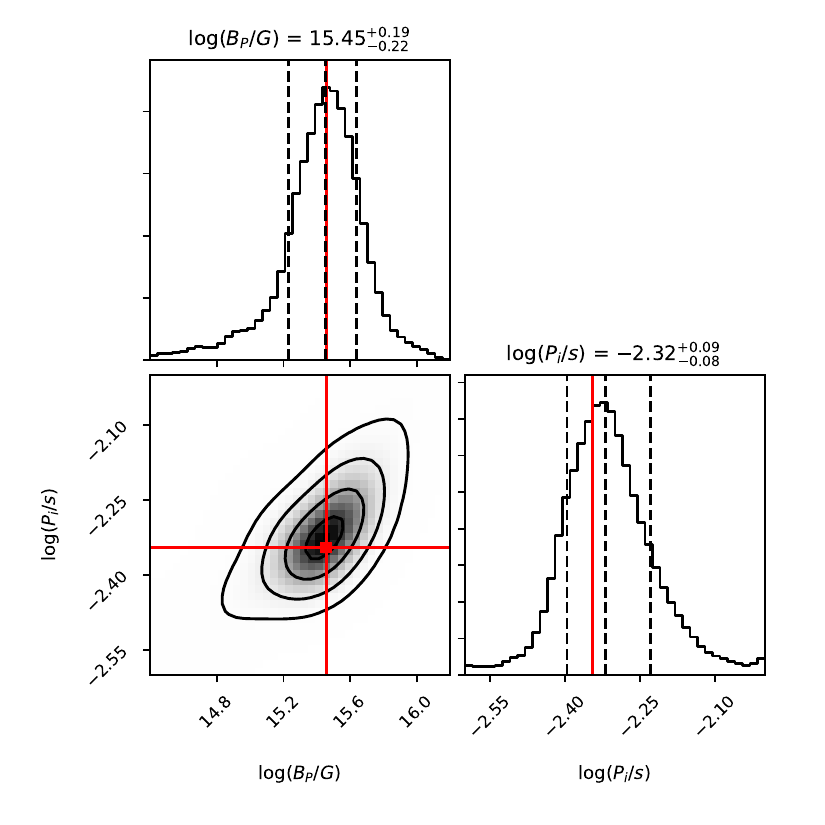}
    \vspace{-0.3cm}
    \caption{Similar to Fig~\ref{fig:Ch4_posterior_free_FXT7}, but for FXT~10.}
    \label{fig:Ch4_posterior_free_FXT10}
\end{figure*}

\begin{figure*}
    \hspace{-1.5cm}
    \centering
    \includegraphics[scale=0.55]{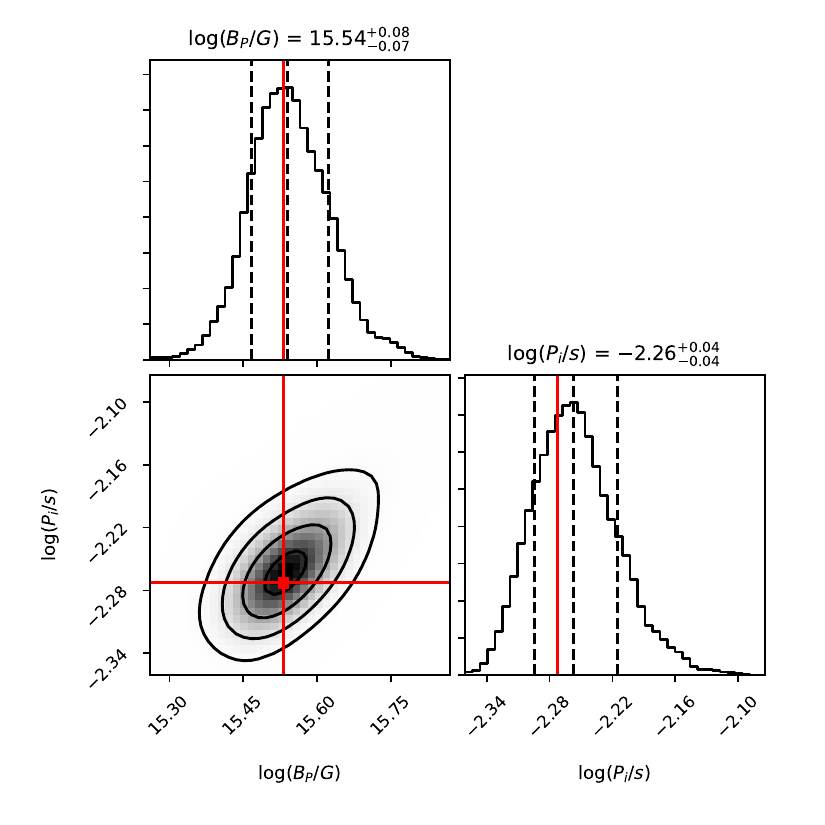}
    \includegraphics[scale=0.55]{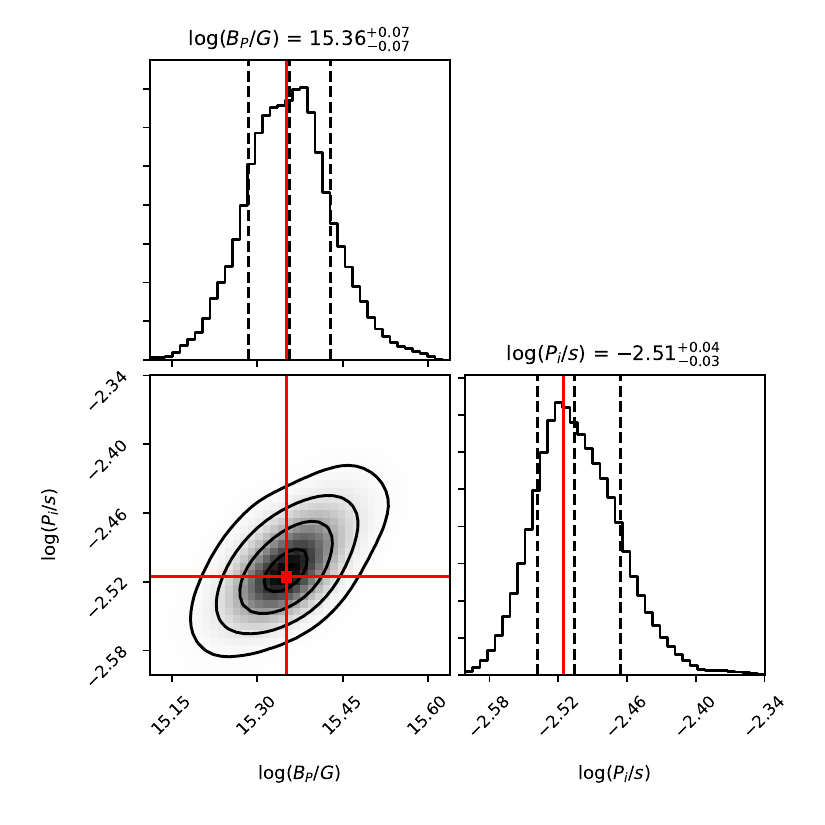}
    \vspace{-0.3cm}
    \caption{Similar to Fig~\ref{fig:Ch4_posterior_free_FXT7}, but for FXT~12.}
    \label{fig:Ch4_posterior_free_FXT12}
\end{figure*}

\begin{figure*}
    \hspace{-1.5cm}
    \centering
    \includegraphics[scale=0.55]{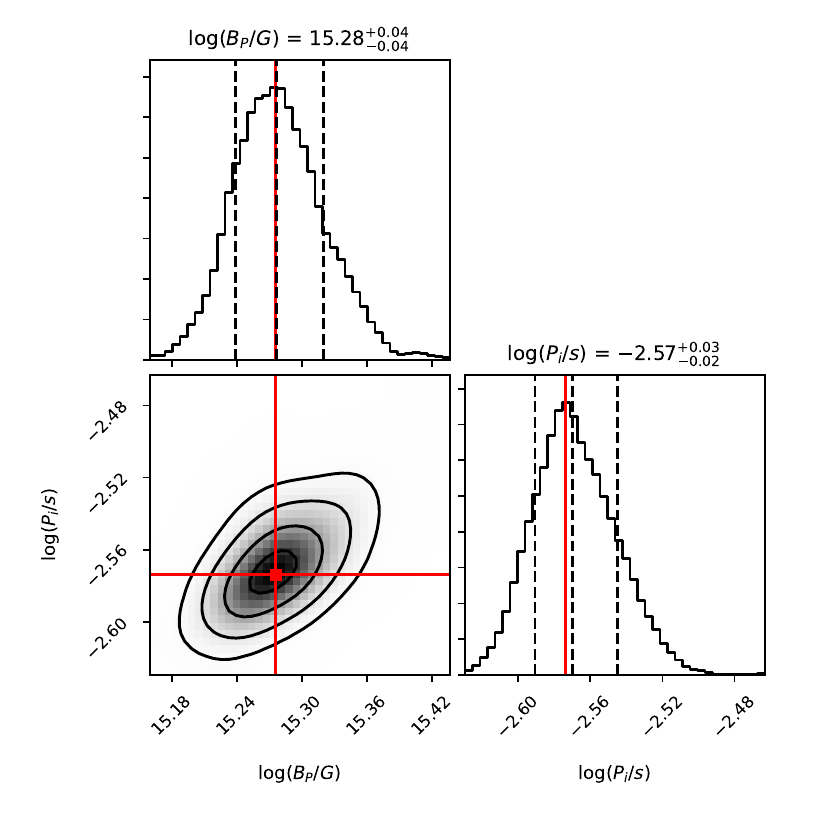}
    \includegraphics[scale=0.55]{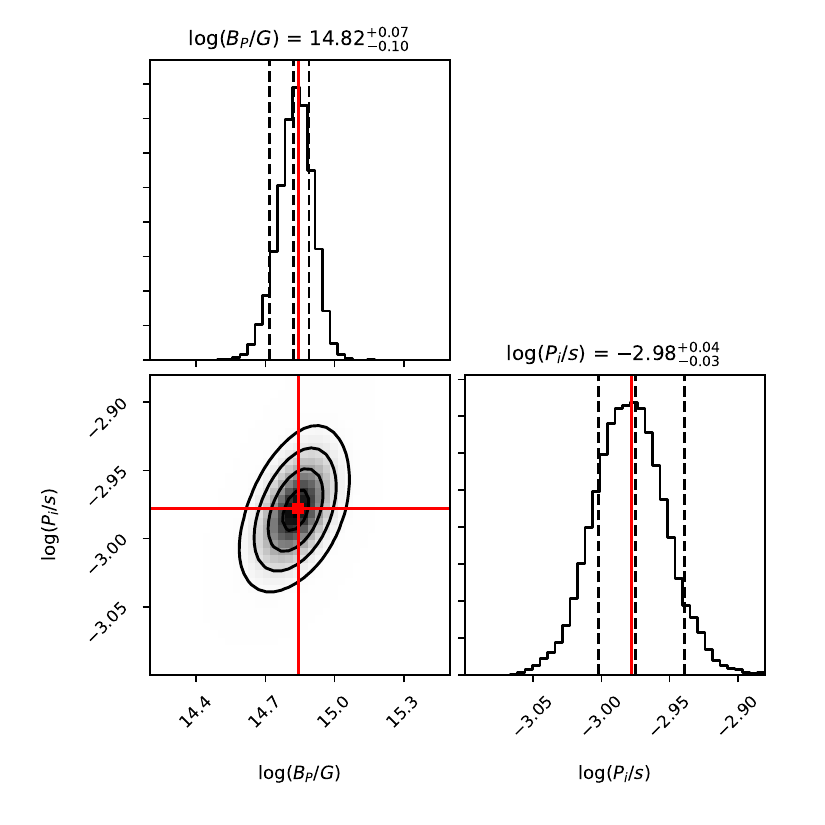}
    \vspace{-0.3cm}
    \caption{MCMC posterior distribution for the free zone fitting result of
FXTs~16 (left panel) and 19 (right panel).}
    \label{fig:Ch4_posterior_free_FXT1619}
\end{figure*}

\begin{figure*}
    \hspace{-1.5cm}
    \centering
    \includegraphics[scale=0.55]{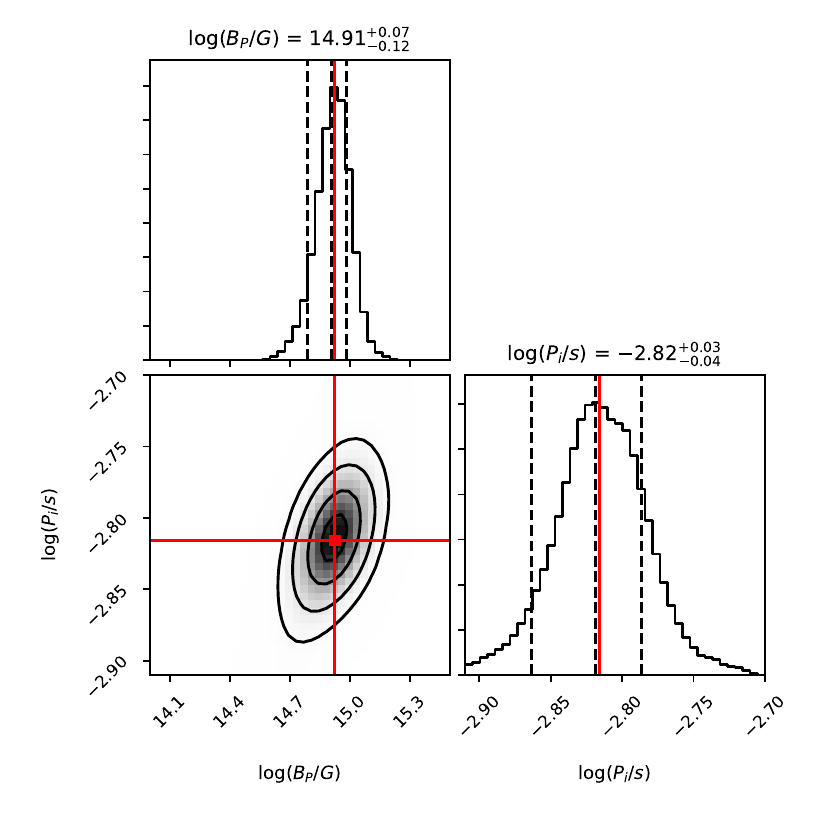}
    \includegraphics[scale=0.55]{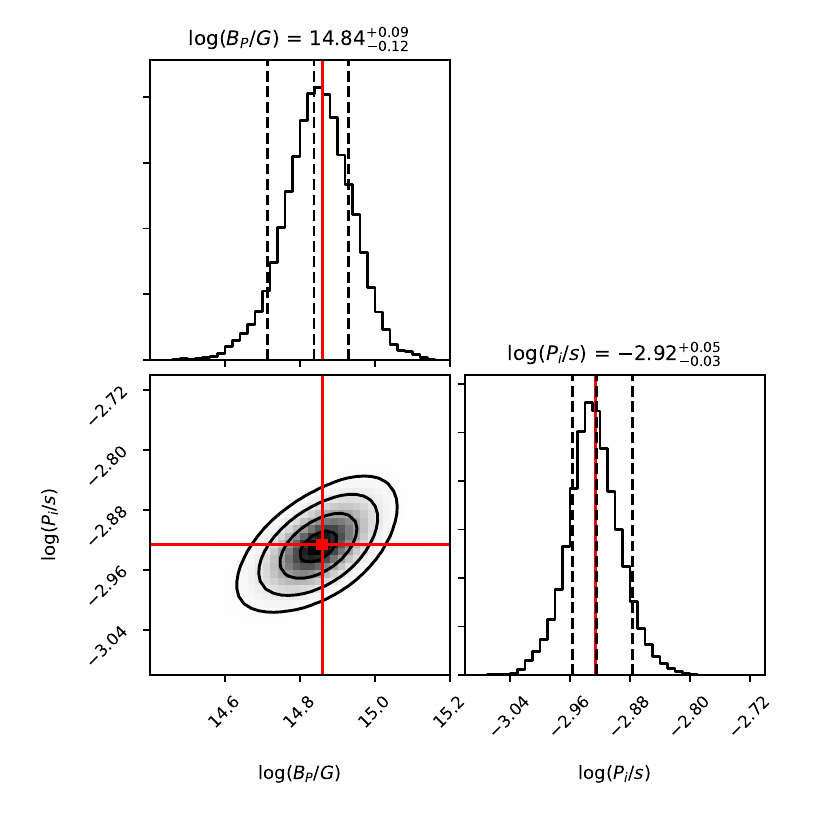}
    \vspace{-0.3cm}
    \caption{MCMC posterior distribution for the free zone fitting result of
FXT~22 assuming $z_{\rm cW}{=}1.04$ (left panel) and $z_{\rm cNE}{=}1.5105$ (right panel).}
    \label{fig:Ch4_posterior_free_FXT22}
\end{figure*}

\begin{figure*}
    \centering
    \hspace{-0.9 cm}
    \includegraphics[scale=0.43]{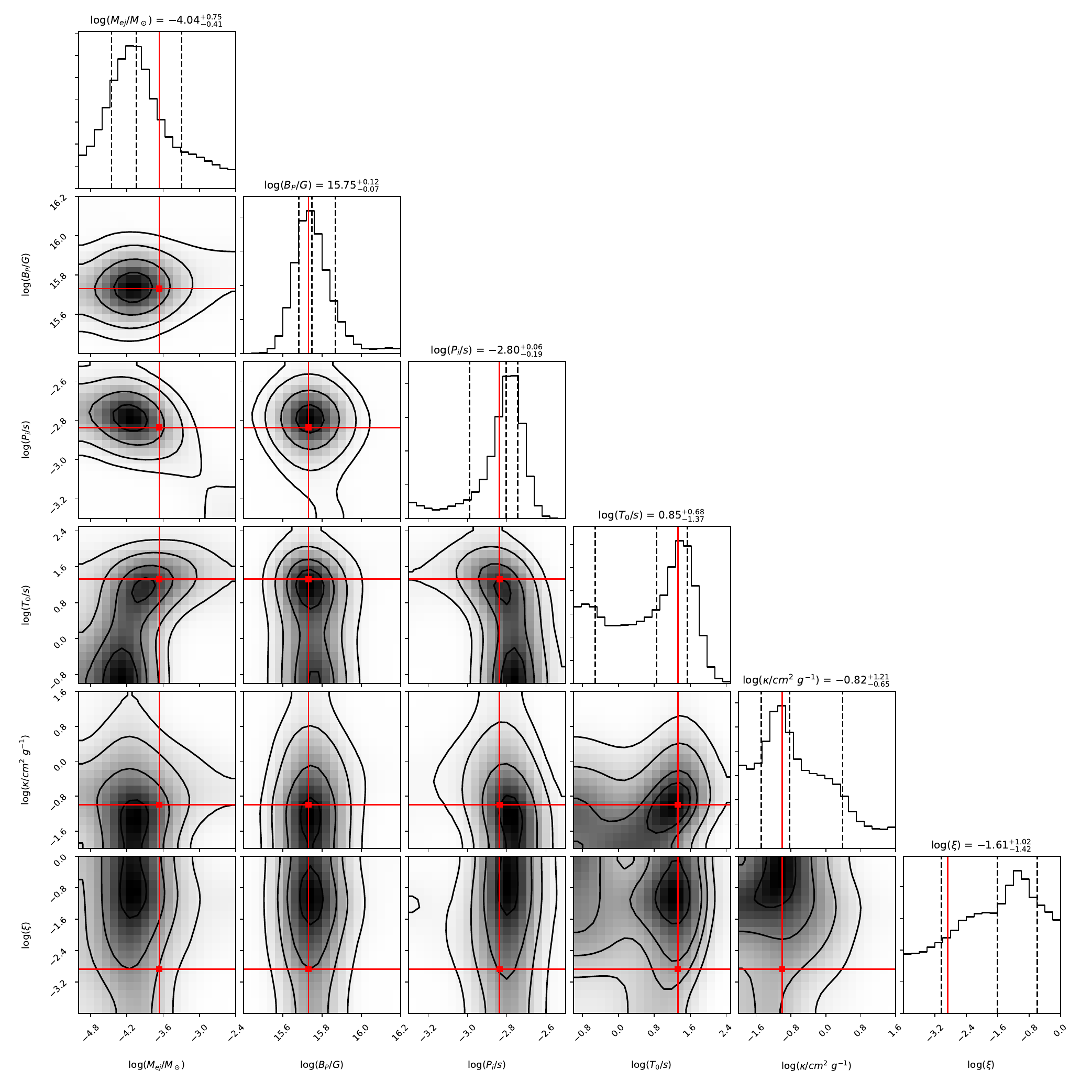}
    \caption{MCMC posterior distribution for the trapped zone fitting result of FXT~14. The dashed black lines show the 16\%, 50\% and 84\% percentile of the MCMC results, while the red lines depict the parameters of the best-fitted model.}
    \label{fig:MCMC_trapped}
\end{figure*}

\begin{figure*}
    \hspace{-1.5cm}
    \centering
    \includegraphics[scale=0.5]{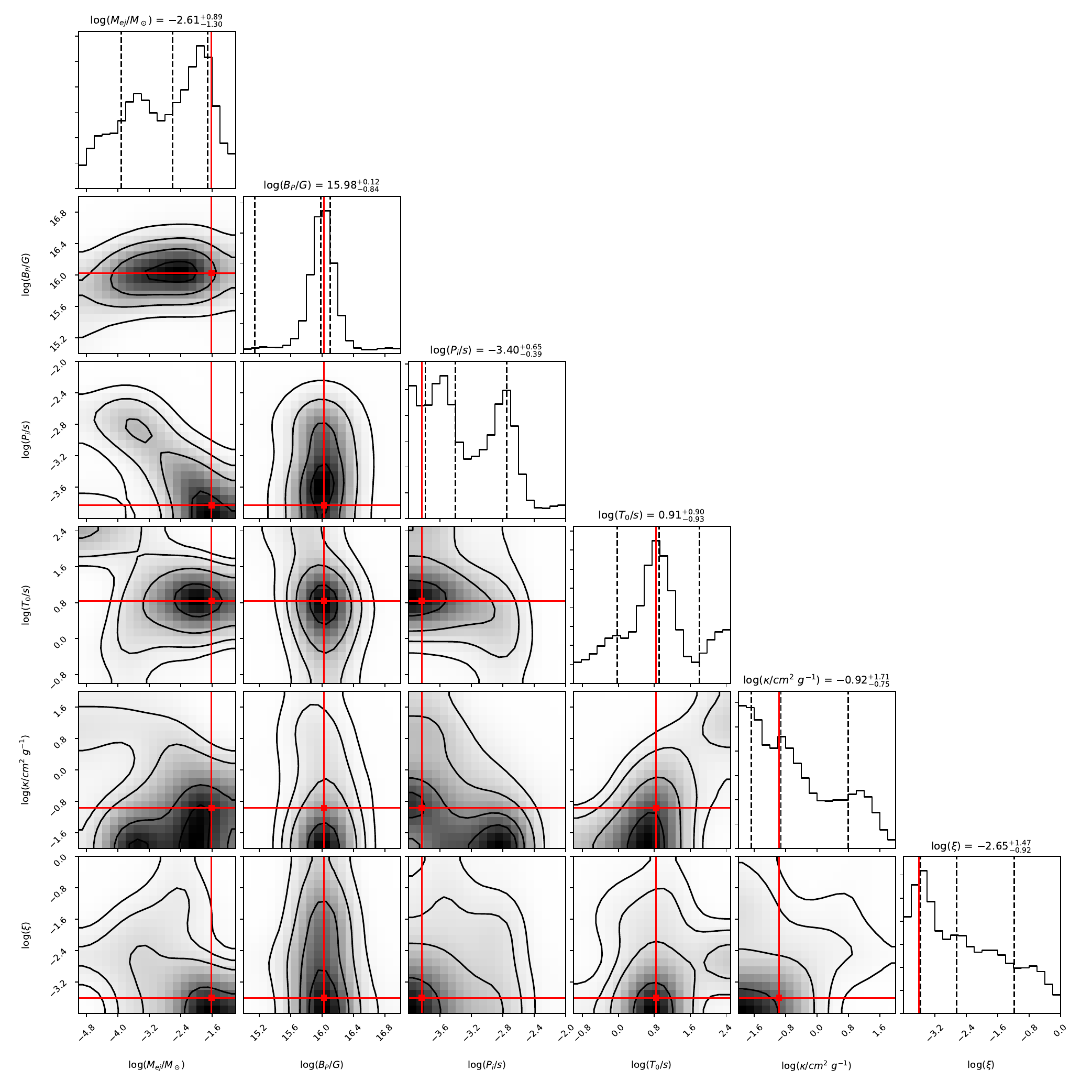}
    \vspace{-0.3cm}
    \caption{Similar to Fig~\ref{fig:MCMC_trapped}, but for 
FXT~20 assuming $z{=}1.0$.}
    \label{fig:Ch4_posterior_trapped_FXT20_1}
\end{figure*}

\begin{figure*}
    \hspace{-1.5cm}
    \centering
    \includegraphics[scale=0.5]{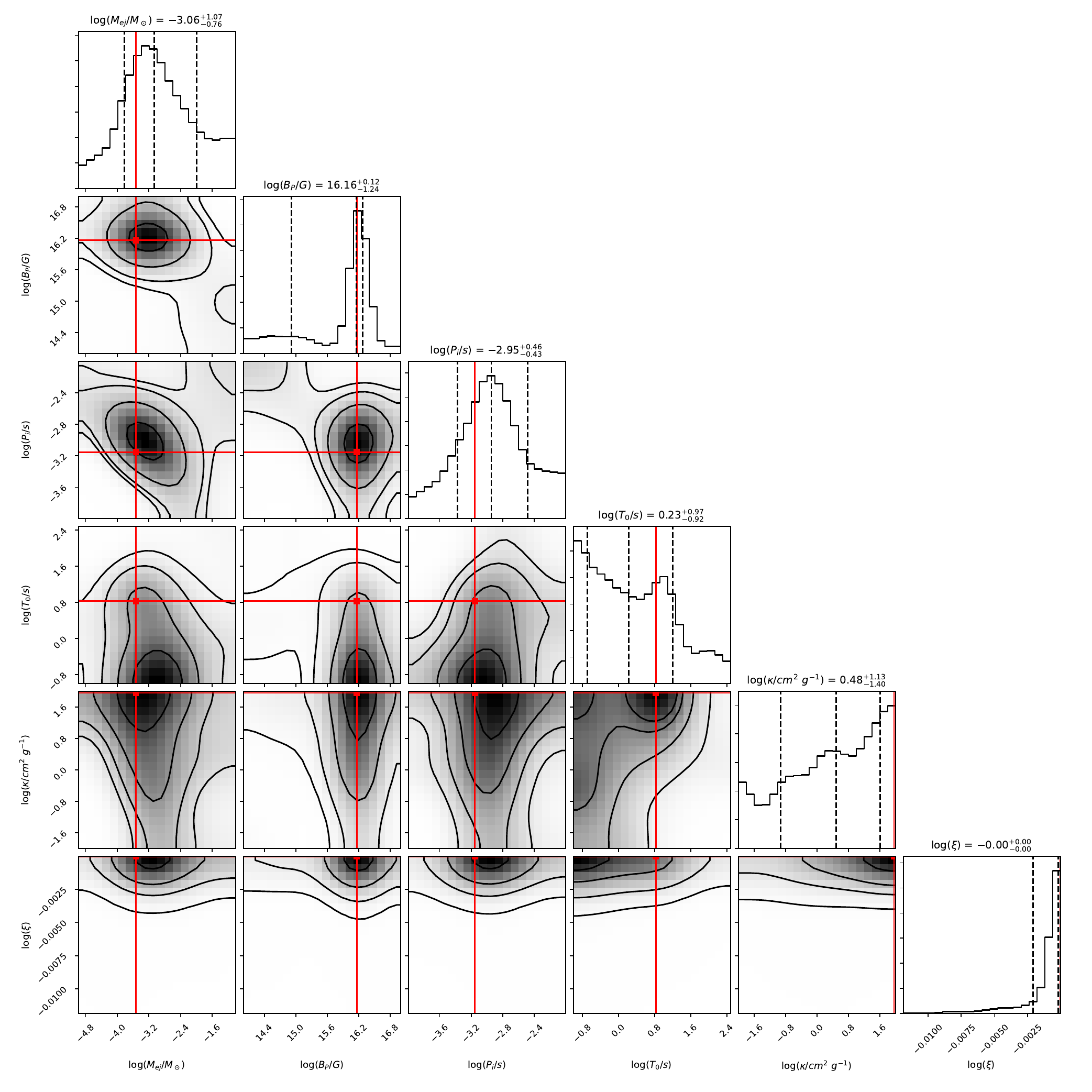}
    \vspace{-0.3cm}
    \caption{Similar to Fig~\ref{fig:MCMC_trapped}, but for FXT~20 assuming $z{=}0.5$.}
    \label{fig:Ch4_posterior_trapped_FXT20_2}
\end{figure*}

\begin{figure*}
    \hspace{-1.5cm}
    \centering
    \includegraphics[scale=0.7]{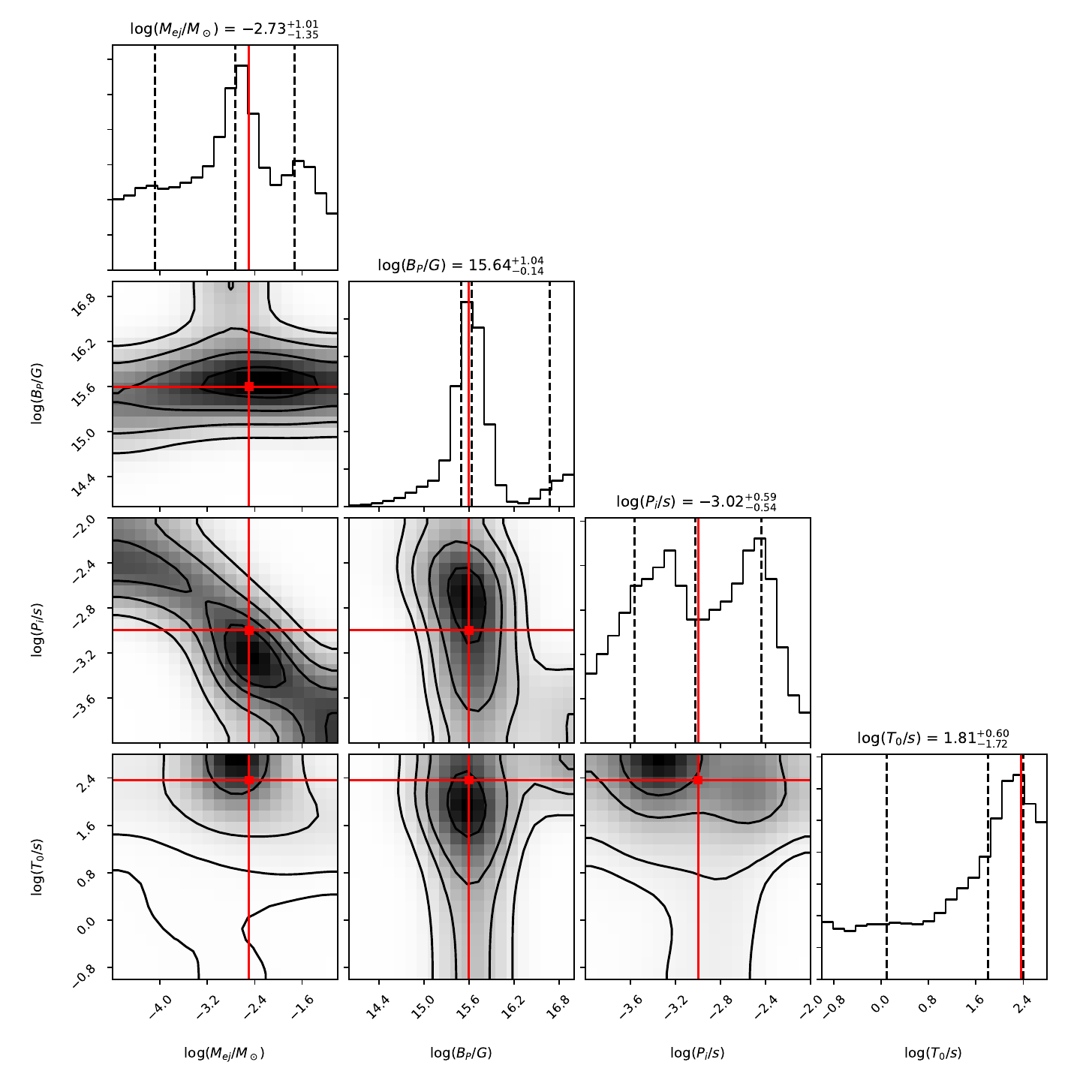}
    \vspace{-0.3cm}
    \caption{Similar to Fig~\ref{fig:MCMC_trapped}, but for 
FXT~15 assuming $z{=}1.0$.}
    \label{fig:Ch4_posterior_trapped_FXT15_1}
\end{figure*}

\begin{figure*}
    \hspace{-1.5cm}
    \centering
    \includegraphics[scale=0.7]{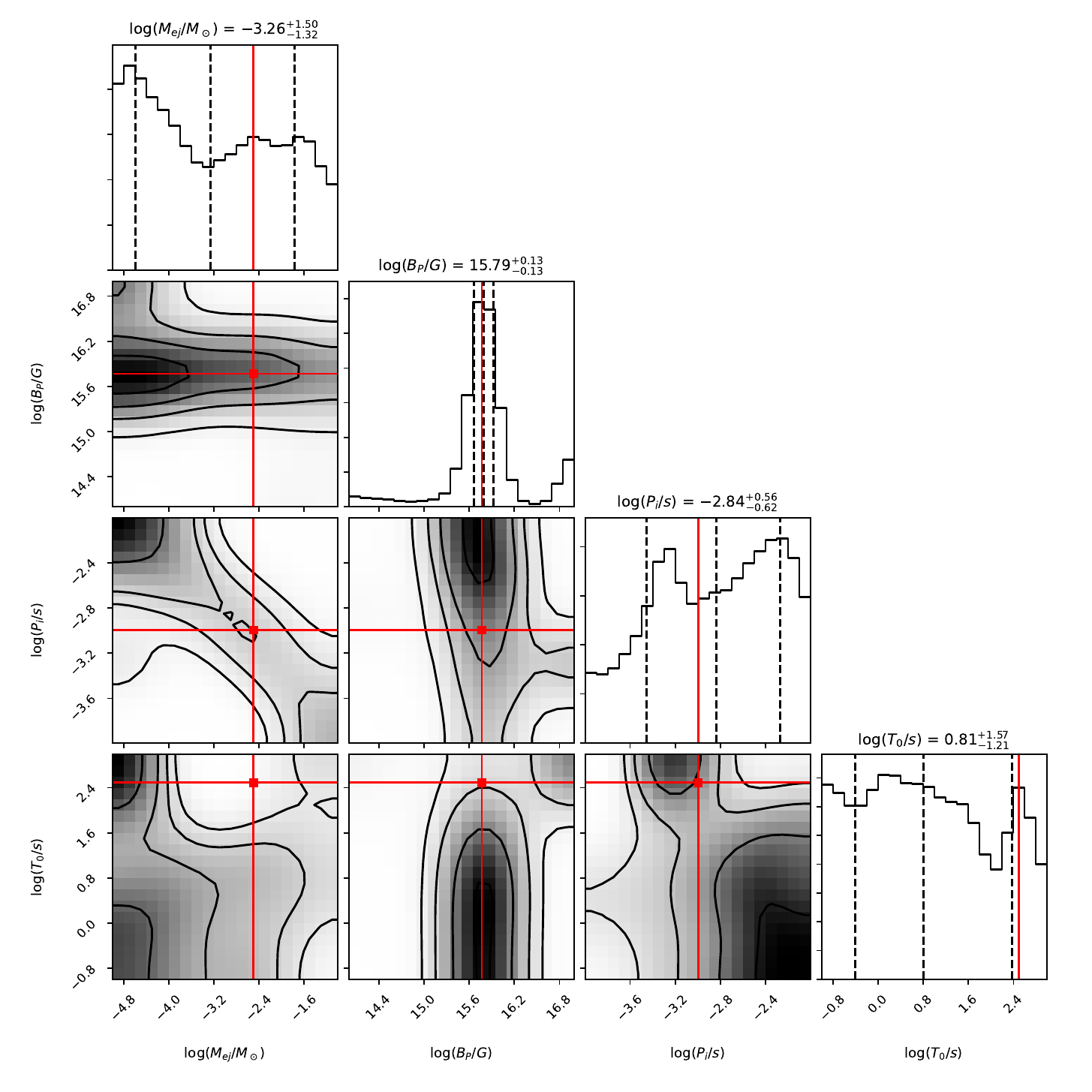}
    \vspace{-0.3cm}
    \caption{Similar to Fig~\ref{fig:MCMC_trapped}, but for 
FXT~15 assuming $z{=}0.5$.}
    \label{fig:Ch4_posterior_trapped_FXT15_2}
\end{figure*}

This appendix shows the posterior distributions (from Fig.~\ref{fig:Ch4_posterior_free_FXT7} to \ref{fig:Ch4_posterior_trapped_FXT15_2}) of the free and trapped model fitting.

\label{lastpage}
\end{document}